\documentclass{article}
\newcommand{\cai}[1]{{#1}}
\usepackage{amsmath,amssymb, pxfonts}
\usepackage{graphicx, color, appendix}
\newtheorem{lem}{Lemma}[section]

\newtheorem{thm}{Theorem}[section]

\newtheorem{ex}{Example}[section]

\title{Asymptotic replication with modified volatility under small transaction costs}
\author{Jiatu Cai$^{\ast}$ and Masaaki Fukasawa$^{\dagger}$
\\
$~~$\\
{\normalsize
$^\ast$ Laboratoire de Probabilit\'es et Mod\`eles Al\'eatoires,} \\ {\normalsize
    Universit\'e Paris Diderot (Paris 7)} \\
$~~$\\
{\normalsize
    $^\dagger$ Department of Mathematics, and} \\
{\normalsize Center for the Study of Finance and Insurance,}
\\ {\normalsize    Osaka University}  \\
{\normalsize 
{fukasawa@math.sci.osaka-u.ac.jp}}
}

\begin{document}
\maketitle
\begin{abstract}
 Dynamic hedging of an European option under a general local volatility model with small \cai{linear} transaction costs is studied.
A continuous control version of Leland's strategy that asymptotically replicates the payoff  is constructed. An associated central limit theorem of hedging  error is proved.
The asymptotic error variance is minimized by an explicit trading strategy.
\end{abstract}

\begin{flushleft}

\vspace*{0.5cm}

\vspace*{0.5cm}

\end{flushleft}

\section{Introduction}
This paper considers  dynamic hedging of an European option under zero interest rate and nonzero transaction costs.
Let $S$ be the price process of the underlying asset of the option and assume
\begin{equation*}
 \mathrm{d}\langle S \rangle_t  = \sigma(S_t,t)^2 \mathrm{d}t
\end{equation*}
on $[0,T]$ with a Borel function $\sigma$ on $(0,\infty) \times [0,T]$. Then for a large class of Borel functions $f$,
\begin{equation*}
 f(S_T) = p(S_0,0) + \int_0^T \partial_sp(S_u,u)\mathrm{d}S_u
\end{equation*}
almost surely, where $p$ is a solution of the partial differential equation (PDE)
\begin{equation*}
 \begin{split}
  &\partial_t p(s,t)+ \frac{1}{2}\sigma(s,t)^2  \partial_s^2 p(s,t) = 0,\\
  & p(s,T) = f(s).
 \end{split}
\end{equation*}
This means that the payoff  $f(S_T)$ is perfectly hedged by the dynamic trading strategy $\partial_sp(S_u,u)$ with initial wealth $p(S_0,0)$ under a hypothetical situation that the strategy incurs no transaction cost. A standard model of transaction costs assumes that the wealth process $\Pi$ associated with a trading strategy $X$ follows
\begin{equation}\label{tcmodel}
 \Pi_t = \Pi_{0-} + \int_0^tX_u\mathrm{d}S_u - 
\kappa \int_{[0,t]} S_u \mathrm{d}\|X\|_u,
\end{equation}
where $\|\cdot\|$ denotes the total variation and $\kappa$ is a positive constant which is supposed to be small.
The model represents that a transaction rebalancing a position from $X_t$ to $X_{t+}$ costs $S_t\Delta X_t$ plus an additional small amount of money which is proportional to $|S_t \Delta X_t|$.
Under this model, it is known that the super replication price coincides with the cost required by the trivial buy-and-hold hedging strategy. In other words, dynamic rebalancing is helpless to hedge European options in the almost sure sense.
See Soner et al.~\cite{Soner} and Corollary 1.6.2 of Kabanov and Safarian~\cite{KS}.
As a result, the super replication price is too expensive in general to be used in practice. One should give up  hedging almost surely and instead, try to reduce hedging error in a distributional sense by a dynamic strategy with a reasonable amount of initial wealth.  

Leland~\cite{Leland} invented such a strategy in an asymptotic framework.
He considered a convex payoff $f$ only but his idea was extended to a general payoff function by Hoggard et al.~\cite{HWW}.
The first trick is to consider the delta hedging strategy with modified volatility.
Let $p^\alpha$ be a solution of the PDE
\begin{equation}\label{pdeal}
 \begin{split}
  &\partial_t p^\alpha(s,t)+ \frac{1}{2}\left(
1 + \mathrm{sgn}(\partial_s^2p^{\alpha}(s,t))\frac{2}{\alpha(s,t)}
\right)\sigma(s,t)^2  \partial_s^2 p^\alpha(s,t) = 0,\\
& p^\alpha(s,T) = f(s),
 \end{split}
\end{equation}
where $\alpha$ is a positive Borel function that controls the modification of volatility.
This is a nonlinear PDE but when $f$ is convex, the solution of the linear PDE
\begin{equation*}
 \begin{split}
  &\partial_t p^\alpha(s,t)+ \frac{1}{2}\left(
1 + \frac{2}{\alpha(s,t)}
\right)\sigma(s,t)^2 \partial_s^2 p^\alpha(s,t) = 0,\\
& p^\alpha(s,T) = f(s)
 \end{split}
\end{equation*}
solves (\ref{pdeal}) 
under a reasonable regularity condition on $f$, $\sigma$ and $\alpha$ which implies  $\partial_s^2 p^\alpha \geq 0$;
see El Karoui et al.~\cite{EJS}. By It$\hat{\text{o}}$'s formula,
\begin{equation*}
 f(S_T) = \Pi^\alpha_0 + \int_0^T X^\alpha_u\mathrm{d}S_u
- \int_0^T \frac{ |\Gamma^\alpha_u|}{\alpha(S_u,u)} \mathrm{d}\langle S \rangle_u,
\end{equation*}
where
\begin{equation}\label{defal}
\Pi^\alpha_t = p^\alpha(S_t,t), \ \ 
 X^\alpha_t = \partial_s p^\alpha(S_t,t), \ \ 
\Gamma^\alpha_t = \partial_s^2 p^\alpha(S_t,t).
\end{equation} 
This means that under no transaction costs, the self-financing strategy $X^\alpha$ with initial wealth $\Pi^\alpha_0$
super-hedges the payoff $f(S_T)$ with surplus
\begin{equation}\label{surp}
 \int_0^T \frac{|\Gamma^\alpha_t|}{\alpha(S_t,t)} \mathrm{d}\langle S \rangle_t \geq 0.
\end{equation}
From an arbitrage argument, we find that 
the initial wealth $\Pi^\alpha_0 = p^\alpha(S_0,0)$ must be larger than $p(S_0,0)$.

The idea is to exploit the surplus (\ref{surp}) to absorb transaction costs. 
Then, the second trick of Leland is to approximate $X^\alpha$ by a good 
sequence of processes of finite variation.
Consider the Black-Scholes model: $\sigma(s,t) = v s$, $v > 0$.
Suppose $f$ to be convex. Take $\alpha$ to be constant.
Leland considered an equidistant discretization of $X^\alpha$; 
define $X^{\alpha, \kappa}$ by
\begin{equation} \label{leland}
X^{\alpha,\kappa}_0 = 0, \ \ 
 X^{\alpha,\kappa}_t = X^\alpha_{jh}, \ \ t \in (jh,(j+1)h], \ \  j = 0,1,2,\dots, 
\end{equation}
where
\begin{equation}\label{leland_h}
 h = \frac{2}{\pi} \frac{\kappa^2 \alpha^2}{v^2}. 
\end{equation}
Set the initial wealth $\Pi^{\alpha,\kappa}_{0-}$, that is, the price of option to be
\begin{equation*}
 \Pi^{\alpha,\kappa}_{0-} = \Pi^\alpha_0 + \kappa S_0 | X^{\alpha}_0|.
\end{equation*}
The associated wealth process $\Pi^{\alpha,\kappa}$ is then
\begin{equation*}
 \Pi^{\alpha,\kappa}_t = \Pi^\alpha_0 + \int_0^t X^{\alpha,\kappa}_u \mathrm{d}S_u - \kappa \sum_{0 < u \leq t} S_u | \Delta X^{\alpha,\kappa}_u|.
\end{equation*}
The magic is that
\begin{equation}\label{magic}
 \begin{split}
&\int_0^T X^{\alpha,\kappa}_t \mathrm{d}S_t
\to \int_0^T X^\alpha_u \mathrm{d}S_u, \\
&\kappa \sum_{0 < u \leq T} S_u | \Delta X^{\alpha,\kappa}_u|
\to
 \frac{1}{\alpha} \int_0^T \Gamma^\alpha_u\mathrm{d}\langle S \rangle_u
=\int_0^T \frac{|\Gamma^\alpha_t|}{\alpha(S_t,t)} \mathrm{d}\langle S \rangle_t
 \end{split}
\end{equation}
as $\kappa \to 0$ with rate  $\kappa$.
As a result, the terminal wealth $\Pi^{\alpha,\kappa}_T$ is close to  $f(S_T)$ when $\kappa$ is small as in reality.
In this sense, the self-financing strategy $X^{\alpha,\kappa}$ is an asymptotic replication strategy with rate $\kappa$.
The way how to discretize $X^\alpha$ is essential.
The first convergence of (\ref{magic}) holds in general as transactions are more and more frequent. On the other hand, if they are too frequent, then the total amount of transaction costs exceeds the surplus (\ref{surp})  and the second convergence of (\ref{magic}) fails to hold.
Therefore the frequency (\ref{leland_h}) results from a delicate balance.
Given $\kappa$, the value of $h$ can be very small if $\alpha$ is very small, which is the case  that the pricing volatility is much enlarged to make the option price close to the super replication price.

\begin{figure}\label{fig1}
\includegraphics[width=6cm]{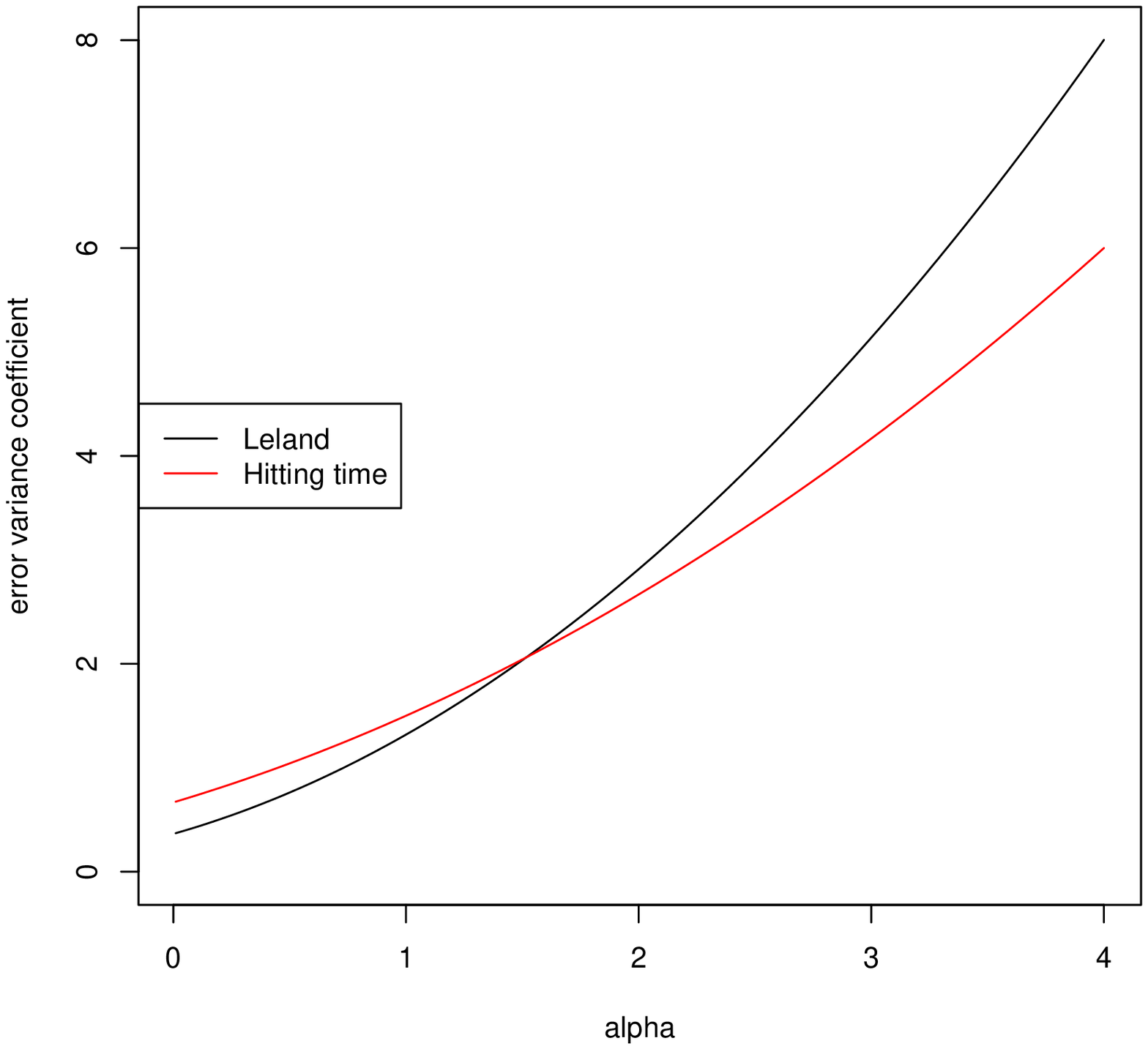}
\includegraphics[width=6cm]{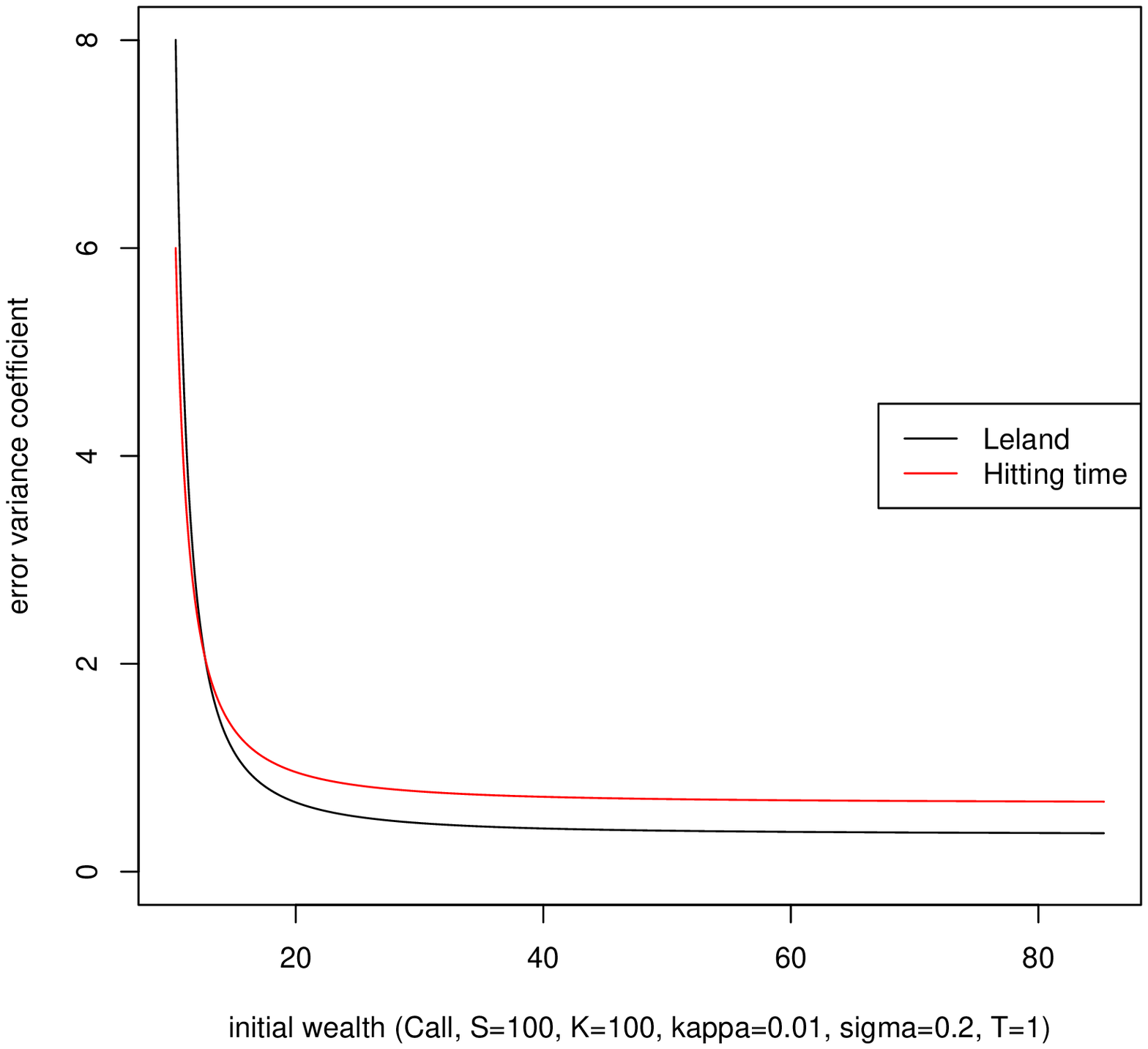}
\caption{Comparison between $\eta_L(\alpha)$ and $\eta_F(\alpha)$.}
\end{figure}

Naturally we expect that a strategy with smaller $\alpha$ that is more costly results in a smaller hedging error. 
In fact it is known for convex payoff functions under the Black-Scholes model that
\begin{equation*}
 \kappa^{-1}(\Pi^\alpha-\Pi^{\alpha,\kappa}) \to W_Q, \ \ 
Q = \eta_L(\alpha)  \int_0^\cdot  |\Gamma^\alpha_u S_u|^2 \mathrm{d}\langle S \rangle_u
\end{equation*}
stably in law on $D[0,T]$ as $\kappa \to 0$, where $W_Q$ is the time-changed process with respect to $Q$ of a standard Brownian motion $W$ which is independent of $S$ and 
\begin{equation*}
 \eta_L(\alpha) = \frac{1}{\pi} \alpha^2 + \frac{2}{\pi} \alpha + 1- \frac{2}{\pi},
\end{equation*}
which is an increasing function of $\alpha$ as shown in the left of Figure~\ref{fig1}.
See Denis and Kabanov~\cite{Kabanov} or Chapter~1 of Kabanov and Safarian~\cite{KS}.
In particular, the asymptotic distribution of $(f(S_T)-\Pi^{\alpha,\kappa}_T)/\kappa$ is mixed normal  with mean zero and variance
\begin{equation*}
Q_T =  \eta_L(\alpha) \int_0^T  |\Gamma^\alpha_u S_u|^2  \mathrm{d}\langle S \rangle_u,
\end{equation*}
which gives a valid approximation to the law of the hedging error $f(S_T) - \Pi^{\alpha,\kappa}_T$ for small but nonzero $\kappa$. This is an important element when considering the trade-off between cost (initial wealth) and risk (hedging error) by controlling $\alpha$; see the right of Figure~\ref{fig1}, where $(\Pi^{\alpha,\kappa}_{0-}, \eta_L(\alpha))$ is plotted for $\alpha \in (0,4)$ and $f(s) = (s-100)_+$, $T=1$, $S_0 = 100$, $\kappa = 0.01$ under the Black-Scholes model with drift $0.01$ and volatility $0.2$.

Leland's strategy can be criticized in that it has no optimality property.
The delta strategy with enlarged volatility $X^\alpha$ is attractive from a practical point of view  even if it is not optimal because its computation is done by a routine work of financial practice.
Still, there is no reason to believe that the equidistant discretization of it is the only choice to approximate it by a strategy of finite variation. In fact, it is known that the discretization with respect to a specific sequence of random times
\begin{equation*}
\tau^\kappa_0= 0, \ \ \tau^\kappa_{j+1} = \inf\{ t > \tau^\kappa ; |X^\alpha_t- X^\alpha_{\tau^\kappa_j}| \geq \alpha\kappa S_{\tau^\kappa_j} \Gamma^\alpha_{\tau^\kappa_j}\}
\end{equation*}
gives
\begin{equation*}
 \kappa^{-1}(\Pi^\alpha-\Pi^{\alpha,\kappa}) \to W_{\hat{Q}}, \ \ 
\hat{Q} = \eta_F(\alpha)  \int_0^\cdot  |\Gamma^\alpha_u S_u|^2 \mathrm{d}\langle S \rangle_u
\end{equation*}
stably in law on $D[0,T)$, where
\begin{equation*}
 \eta_F(\alpha) = \frac{(\alpha+2)^2}{6}.
\end{equation*}
See Figure~\ref{fig1}. 
Remark that
 $\eta_F(\alpha) < \eta_L(\alpha)$ if and only if
\begin{equation*}
 \alpha > \frac{6-2\pi + \sqrt{6(18-8\pi + \pi^2)}}{\pi-6} \approx 1.5168.
\end{equation*}
Therefore,
 this hitting time strategy is superior to the equidistant one in terms of asymptotic mean-squared tracking  error in a practical region of $\alpha$
where the original volatility  $\sigma$ makes a reasonable portion of the pricing volatility $\sigma\sqrt{1+2/\alpha}$. 
Besides, the hitting time structure requires less rebalancing times.
See Fukasawa~\cite{F} for the detail. See also Toft~\cite{Toft} and Ahn et al.~\cite{Ahn} for earlier results.
Still, there is no reason to believe the hitting time strategy is the best.

Another direction of extending Leland's original idea that used a constant $\alpha$ is to consider a time-varying $\alpha$. Then, say, a deterministic but non-equidistant partition may be used to approximate $X^\alpha$. See Grannan and Swindle~\cite{GS} or Denis and Kabanov~\cite{Kabanov}. Then the problem is to find an optimal $\alpha$, which will depend on the payoff function $f$.
We will be able to solve this after we find an optimal way to realize Leland's second trick for a given $\alpha$.

The preceding studies have considered only discretization schemes of $X^\alpha$. However, what we need is just an approximating sequence of processes which are of finite variation.
They do not need to be a discretized process of $X^\alpha$.
On the contrary, results on related problems of utility maximization suggest that simple processes are not efficient for approximating $X^\alpha$ under the proportional transaction costs. See e.g., Whalley and Wilmott~\cite{WW}, Barles and Soner~\cite{BS}, and Soner and Touzi~\cite{ST}. The optimal strategies in those framework are singular control strategies.
The aim of this study is to analyze the asymptotic behavior of hedging error when regular and singular control strategies are used to approximate $X^\alpha$, and to find an optimal one in a suitable sense.
Remark that these continuous controls are practically relevant in that they are approximated by impulse controls.
We show there is in fact a continuous control strategy which drastically reduces the hedging error.

\section{Continuous control strategies}
Here we give a rigorous formulation. 
Let $(\Omega, \mathcal{F}, \mathbb{P}, \{\mathcal{F}_t; t \geq 0\})$ be a filtered probability space satisfying the usual assumptions.
Let $T >0$ be a constant which stands
 for the maturity of an European option. Let $f$ be a Borel function on $(0,\infty)$ which stands for the payoff function of the option. We suppose the underlying asset price process $S$ of the option to be positive and continuous on $[0,T]$ and to follow
\begin{equation*}
 \mathrm{d} S_t = \Theta_t \mathrm{d}t +  \sigma(S_t,t)\mathrm{d}B_t
\end{equation*}
on $[0,T]$, where $\Theta$ is an $\{\mathcal{F}_t\}$-adapted locally bounded process,
$B$ is an $\{\mathcal{F}_t\}$-standard Brownian motion and
$\sigma$ is a positive $C^{2,1}$ functions on $(0,\infty) \times [0,T]$.
For an $\{\mathcal{F}_t\}$-predictable process $X$, taking it as a trading strategy, the associated wealth process $\Pi$ is defined by
\begin{equation}\label{tcmodel2}
 \Pi_t = \Pi_{0-} + \int_0^tX_u\mathrm{d}S_u - 
\kappa \int_{[0,t]}\lambda(S_u,u) \mathrm{d}\|X\|_u,
\end{equation}
which generalizes (\ref{tcmodel}), where $\lambda$ is a nonnegative $C^{2,1}$ function on $(0,\infty) \times [0,T]$.
The constant $\kappa > 0$ appeared in (\ref{tcmodel}) and (\ref{tcmodel2}) represents transaction cost coefficient that is considered to be small in reality. 
We will study the asymptotic behavior of hedging as $\kappa \to 0$, which serves as a valid approximation to the hedging behavior 
when $\kappa$ is sufficiently small. 

Denote by $\mathcal{A}$ the set of $C^{2,1}$ functions $\varphi$ on
$(0,\infty) \times [0,T)$ such that for
each $i \in \{0,1,2\}$ and $j\in \{0,1\}$, 
$\partial_s^i \partial_t^j\varphi(S_t,t)$
converges almost surely as $t \to T$.
Let $\alpha$ be a positive $C^{2,1}$ function on $(0,\infty) \times [0,T]$ which is  so regular that the PDE (\ref{pdeal}) admits a solution  
$p^\alpha$ which is continuous on $(0,\infty) \times [0,T]$ and for
each $i \in \{0,1,2\}$ and $j\in \{0,1\}$, 
$\partial_s^i \partial_t^j p^{\alpha} \in \mathcal{A}$.
Further, we assume  $\partial_s^2p^\alpha$ is nondegenerate in the sense that
\begin{equation}\label{nondeg}
 \int_0^T 1_{\{ |\partial_s^2p^\alpha(S_t,t)| =0\}} \mathrm{d}t=0
\end{equation}
almost surely.
This holds if, say, $|\partial_s^2p^\alpha(s,t)| > 0$ for all $(s,t) \in (0,\infty) \times [0,T)$ by the bounded convergence theorem. The simplest example that satisfies all of the above assumptions is 
Leland's original framework:
the Black-Scholes model for $S$,  the call or put payoff for $f$,
$\lambda(s,t) =s$
 and a positive constant for $\alpha$.

Define $\Pi^\alpha, X^\alpha$ and $ \Gamma^\alpha$ by (\ref{defal}). Recall  $f(S_T) =\Pi^\alpha_T$.
We regard $\Pi^\alpha$  as a benchmark portfolio value and consider the tracking error
\begin{equation} \label{trer}
 \Pi^\alpha_t- \Pi_t = \int_0^t (X_u^\alpha-X_u)\mathrm{d}S_u + \kappa \int_{(0,t]} \lambda(S_u,u)\mathrm{d}\|X\|_u - \int_0^t\frac{|\Gamma^\alpha_u|}{\alpha(S_u,u)} \mathrm{d}\langle S \rangle_u 
\end{equation}
for a trading strategy $X$. Here we set the initial wealth as 
\begin{equation}\label{iniw}
 \Pi_{0-} = \Pi^\alpha_0 + \kappa\lambda(S_0,0) |\Delta X_0|.
\end{equation}
In order to keep the tracking error finite, $X$ must be of finite variation. A simple predictable process is of finite variation. All the preceding studies for Leland-type strategies have considered discretized processes of $X^\alpha$ that are simple predictable processes.
In this study, we consider a class of continuous processes of finite variation.
In order to make the tracking error small, a reasonable control would be based on the deviation between $X^\alpha$ and $X$.
Let $Z^\kappa = (X^\alpha - X)/\kappa$.
We will consider $X$ of the form
\begin{equation} \label{gamma}
 \mathrm{d}X_t = \frac{1}{\kappa}
\mathrm{sgn}(Z^\kappa)c(|Z^\kappa_t|,S_t,t)\nu(S_t,t)^2\mathrm{d}t 
- \kappa \mathrm{d}L^\kappa_t + \kappa \mathrm{d}R^\kappa_t, 
\ \ X_{0+} = X^\alpha_0
\end{equation}
where $c$ is a nonnegative Borel function,
\begin{equation*}
 \nu(s,t) = \sigma(s,t) \partial_s^2p^\alpha(s,t)
\end{equation*}
and  $L^\kappa$ and $R^\kappa$ are nondecreasing processes such that
\begin{equation}\label{singp}
 L^\kappa_t = \int_0^t 1_{\{Z^\kappa_u = - b(S_u,u)\}} \mathrm{d}L^\kappa_u, \ \ 
 R^\kappa_t = \int_0^t 1_{\{Z^\kappa_u =  b(S_u,u)\}} \mathrm{d}R^\kappa_u, \ \ |Z^\kappa_t| \leq b(S_t,t)
\end{equation} 
on $[0,T]$ for a positive Borel function $b$.
The idea is to introduce a regular control part which pushes $X$ up or down if $Z^\kappa$ is positive or negative respectively and
a singular control part which keeps  $Z^\kappa$ within a stochastic interval.
The function $\nu$ is introduced for notational convenience in the sequel.

The existence of such $L^\kappa$ and $R^\kappa$ follows from that of a solution of a Skorokhod-type equation. 
Denote by $\mathcal{B}$ the set of the positive functions $b$ on $(0,\infty) \times [0,T)$ such that both $b$ and $1/b$ belong to  $\mathcal{A}$.
For $b \in \mathcal{B}$, denote
by $\mathcal{C}_b$  the set of nonnegative and piecewise $C^{0,2,1}$ functions $c$ on 
\begin{equation*}
\mathcal{D}_b:= \{(x,s,t) \in \mathbb{R} \times (0,\infty) \times [0,T) \ ; \  |x| \leq b(s,t)\}
\end{equation*}
such that 
\begin{enumerate}
\item 
for all $(s,t)$,  $c(\cdot, s,t)$ are even:
\begin{equation*}
c(x,s,t) = c(-x,s,t),
\end{equation*}
\item
for all $x$,  $c(x,\cdot)$ are $C^{2,1}$  and
\begin{equation}\label{marg}
\sup\{ |\partial_s^i \partial_t^jc(x,S_t,t)| ; \ 
t \in [0,T), x \in [-b(S_t,t),b(S_t,t)] \}
< \infty
\end{equation}
almost surely  for each $i \in \{0,1,2\}$ and $j \in \{0,1\}$.
\item
 for any compact set $A \subset \mathcal{D}_b$,  there exists $K > 0$ such that
\begin{equation}\label{lip}
(x-y)(-\mathrm{sgn}(x)c(x,s,t) + \mathrm{sgn}(y)c(y,s,t)) \leq K|x-y|^2
\end{equation}
for all $(x,s,t), (y,s,t) \in A$.  
\end{enumerate}
For $b \in \mathcal{B}$ and $c \in \mathcal{C}_b$, by a fixed point argument thanks to the one-sided Lipschitz condition (\ref{lip}) (see e.g., Tanaka~\cite{Tanaka}), we can show that there exists a unique solution 
$(Z^\kappa, L^\kappa ,R^\kappa)$ of a Skorokhod-type equation
\begin{equation}\label{sko}
\mathrm{d}Z^\kappa_t  =  
\frac{1}{\kappa} \mathrm{d}X^\alpha_t
-
\frac{1}{\kappa^2}\mathrm{sgn}(Z^\kappa_t)c(Z^\kappa_t,S_t,t)
\nu(S_t,t)^2 \mathrm{d}t + \mathrm{d}L^\kappa_t - \mathrm{d}R^\kappa_t,  \ \ Z^\kappa_0 = 0
\end{equation}
with (\ref{singp}) on $[0,T]$.  Therefore the strategy (\ref{gamma}) is well-defined for each $b \in \mathcal{B}$ and $c \in \mathcal{C}_b$. The total variation of $X$ is then given by
\begin{equation*}
 \mathrm{d}\|X\|_t = 
\frac{1}{\kappa} c(Z^\kappa_t,S_t,t)\nu(S_t,t)^2\mathrm{d}t + \kappa [\mathrm{d}L^\kappa_t + \mathrm{d}R^\kappa_t].
\end{equation*}

Denote by $\Pi^{b,c,\kappa}$ the associated wealth process 
with initial capital (\ref{iniw}).
Then, from (\ref{trer}), the associated tracking error $\mathcal{E}^{b,c,\kappa}$ is given by
\begin{equation*}
\begin{split}
 \mathcal{E}^{b,c,\kappa}_t 
= & \Pi^\alpha_t - \Pi^{b,c,\kappa}_t \\
= &\kappa \int_0^t Z^\kappa_u \mathrm{d}S_u + 
\int_0^t \lambda(S_u,u)c(Z^\kappa_u,S_u,u)\nu(S_u,u)^2\mathrm{d}u 
\\ & + \kappa^2 \int_0^t \lambda(S_u,u)[\mathrm{d}L^\kappa_u + \mathrm{d}R^\kappa_u]
- \int_0^t \frac{|\Gamma^\alpha_u|}{\alpha(S_u,u)} \mathrm{d}\langle S \rangle_u.
\end{split}
\end{equation*}
So far we can freely choose $b$ and $c$. 
The question is whether the tracking error converges to $0$ as $\kappa \to 0$ with a good rate. 
If the answer is positive for a certain class of $b$ and $c$, then the next question is which combination of $b$ and $c$ is optimal. We will answer these \cai{questions} in the following sections.

\section{Limit theorem of hedging error}

The aim of this section is to prove the following new result.

\begin{thm}\label{main}
Let $b \in \mathcal{B}$ and $c \in \mathcal{C}_b$. %Suppose a solution of (\ref{sko}) exists. 
Then,
\begin{equation*}
\kappa^{-1}\left(
 \mathcal{E}^{b,c,\kappa} -  \int_0^\cdot \delta(S_t,t)\mathrm{d}\langle S \rangle_t
\right)
\end{equation*}
converges stably in law on $C[0,T]$ to a time-changed Brownian motion $W_Q$ as $\kappa \to 0$, where
$W$ is a standard Brownian motion independent of $\mathcal{F}$ and
\begin{equation}\label{funs}
 \begin{split}
&Q = \int_0^\cdot \eta_{b,c}(S_t,t)\mathrm{d}\langle S \rangle_t, \\
& \eta_{b,c}(s,t) = \frac{2}{a(s,t)}\int_0^{b(s,t)}(x-\lambda(s,t)\partial_s^2p^\alpha(s,t)h(x,s,t))^2g(x,s,t)\mathrm{d}x, \\
&g(x,s,t) = \exp\left\{ - 2 \int_0^{|x|}c(y,s,t) \mathrm{d}y \right\},\\
&a(s,t) = 2 \int_0^{b(s,t)} g(x,s,t) \mathrm{d}x, \\ 
& h(x,s,t) = \frac{2 \mathrm{sgn}(x)}{g(x,s,t)}\int_0^{|x|} \left(c(z,s,t) - \frac{1}{a(s,t)}\right)g(z,s,t)\mathrm{d}z, \\
& \delta(s,t) = \frac{\lambda(s,t)|\partial_s^2p^\alpha(s,t)|^2}{a(s,t)} - \frac{|\partial_s^2p^\alpha(s,t)|}{\alpha(s,t)}.
 \end{split}
\end{equation}
\end{thm}
In particular, 
by taking $b \in \mathcal{B}$ and $c \in \mathcal{C}_b$ such that $a \geq \alpha \lambda |\partial_s^2p^\alpha|$, 
we have an asymptotic super replication strategy:
\begin{equation*}
\Pi^{b,c,\kappa}_T \to f(S_T) - \int_0^T \delta(S_t,t)\mathrm{d}\langle S \rangle_t \geq f(S_T)
\end{equation*}
as $\kappa \to 0$. This is always possible; say, let
\begin{equation}\label{pures}
 b(s,t) = \frac{1}{2} \alpha(s,t) \lambda(s,t) |\partial_s^2p^\alpha(s,t)| + \epsilon, \ \ c(x,s,t)=0
\end{equation} 
with $\epsilon > 0$. Then we have $a > \alpha \lambda |\partial_s^2p^\alpha|$. 
If $|\partial_s^2p^\alpha| > 0$ on $(0,\infty) \times [0,T]$, one can take $\epsilon = 0$ to have
$a = \alpha \lambda |\partial_s^2p^\alpha|$. 
\cai{Before starting the proof, we give an example in comparison to the existing results.}

\begin{ex}%[Black-Scholes Model with Convex Payoff]
\upshape
Consider the Black-Scholes model: $\sigma(s,t) = vs$ and $\lambda(s,t) = s$. 
Let $f$ be convex and $\alpha$ be constant. 
Then $p^\alpha$ is the Black-Scholes price with enlarged volatility $v\sqrt{1+\alpha/2} > v$. As explained in Introduction, the original Leland strategy uses the equidistant discretization of $X^\alpha$ with (\ref{leland_h}). The renormalized tracking error at time $t \in [0,T]$ converges in law to the mixed normal distribution with mean $0$ and variance
\begin{equation*}
 \eta_L(\alpha)  \int_0^t |S_t\Gamma^\alpha_t|^2\mathrm{d}\langle S \rangle_t.
\end{equation*}
The use of the hitting times changes the limit variance to
\begin{equation*}
 \eta_F(\alpha)  \int_0^t |S_t\Gamma^\alpha_t|^2\mathrm{d}\langle S \rangle_t
\end{equation*}
without changing the initial  wealth $p^\alpha(S_0,0) + \kappa S_0 |X^\alpha_0|$.
Now, let us consider the simplest control strategy in our framework for the same constant $\alpha$. 
Let $b \in \mathcal{B}$ and  $c = 0 \in \mathcal{C}_b$. Then, we have
\begin{equation*}
g(z,s,t) = 1,\ \
 a(s,t) = 2b(s,t), \ \
h(x,s,t) = -x/b(s,t)
\end{equation*}
by definition and so,
\begin{equation*}
 \delta(s,t) = \frac{s|\partial_s^2p^\alpha(s,t)|^2}{2b}- \frac{|\partial_s^2p^\alpha(s,t)|}{\alpha}, \ \
 \eta_{b,c}(s,t) = \frac{1}{3}\left(b(s,t) + s\partial_s^2p^\alpha(s,t)\right)^2.
\end{equation*}
With the same initial  wealth $p^\alpha(S_0,0) + \kappa S_0 |X^\alpha_0|$ as before,
the hedging error converges to $0$ with rate $\kappa$ by taking $b$ as (\ref{pures}) with $\epsilon = 0$.
In this case, the limit variance of the renormalized hedging error is
\begin{equation*}
\frac{(\alpha+2)^2}{12}  \int_0^t |S_t\Gamma^\alpha_t|^2\mathrm{d}\langle S \rangle_t.
\end{equation*}
It is easy to see
\begin{equation*}
  \frac{(\alpha+2)^2}{12} < \min\{\eta_L(\alpha), \eta_F(\alpha)\}
\end{equation*}
for all $\alpha > 0$, which means that the simplest singular control is already superior to the existing ones in terms of the asymptotic mean-squared error.
 Further, our result is valid for a general local volatility model.
These superiorities remain to hold even if we consider a time-varying $\alpha$ as in Denis and Kabanov~\cite{Kabanov}.
We will minimize $\eta_{b,c}$ in the next section.
\end{ex}

The proof of Theorem~\ref{main} utilizes a homogenization technique of two-scale stochastic differential equations and the theory of scale function and speed measure for one-dimensional ergodic diffusions. In fact the function $g$ defined in (\ref{funs}) is the speed measure for the corresponding diffusion. See Skorokhod~\cite{Sko}, Papavasiliou et al.~\cite{PPS} and Glynn and Wang~\cite{GW} for more general results. We start with the following lemma.

\begin{lem}\label{centered0}
Let $\psi$ be a piecewise $C^{0,2,1}$ function on $\mathcal{D}_b$ such that
\begin{enumerate}
\item for each $(s,t)$,
\begin{equation*}
 \int_{-b(s,t)}^{b(s,t)} \psi(x,s,t)g(x,s,t)\mathrm{d}x = 0
\end{equation*}
and
\item for each $x$, $\psi(x,\cdot)$ is $C^{2,1}$  and
\begin{equation}\label{marg}
\sup\{ |\partial_s^i \partial_t^j\psi(x,S_t,t)| ; \ 
t \in [0,T), x \in [-b(S_t,t),b(S_t,t)] \}
< \infty
\end{equation}
almost surely  for each $i \in \{0,1,2\}$ and $j \in \{0,1\}$.
\end{enumerate}
Then,      
\begin{equation*}
 \sup_{t \in [0,T]} \left|\int_0^t \psi(Z^\kappa_u,S_u,u)\nu(S_u,u)^2\mathrm{d}u \right| \to 0 
\end{equation*} 
in probability as $\kappa \to \infty$.
\end{lem}
{\it Proof: } Let
\begin{equation*}
\begin{split}
& \Psi(x,s,t) = \int_0^x \psi_1(z,s,t)\mathrm{d}z, \\
& \psi_1(z,s,t) = \frac{2}{g(z,s,t)} \int_{-b(s,t)}^z \psi(x,s,t)g(x,s,t)\mathrm{d}x.
\end{split}
\end{equation*}
Then, $\Psi$ is a $C^{1,2,1}$ and piecewise $C^{2,2,1}$ function and 
\begin{equation}\label{marpsi}
 \psi_1(b(s,t),s,t) =  \psi_1(-b(s,t),s,t) = 0
\end{equation}
by the assumption. Note also that
\begin{equation}\label{genpsi}
 -\mathrm{sgn}(z)c(z,s,t)\psi_1(z,s,t) + \frac{1}{2}\partial_z\psi_1(z,s,t) = \psi(z,s,t).
\end{equation}
By a generalized It$\hat{\text{o}}$ formula of Peskir~\cite{Peskir},
\begin{equation*}
 \begin{split}
  \Psi(Z^\kappa_t,S_t,t) =& \Psi(Z^\kappa_0,S_0,0) + \int_0^t \psi_1(Z^\kappa_u,S_u,u)\mathrm{d}Z^\kappa_u + \int_0^t \partial_s\Psi(Z^\kappa_u,S_u,u)\mathrm{d}S_u \\ 
&+  \int_0^t \partial_t\Psi(Z^\kappa_u,S_u,u)\mathrm{d}u + \frac{1}{2} \int_0^t \partial_s^2 \Psi(Z^\kappa_u,S_u,u)\mathrm{d}\langle S \rangle_u \\ & + \int_0^t \partial_s\psi_1(Z^\kappa_u,S_u,u)\mathrm{d}\langle Z^\kappa, S \rangle_u + 
\frac{1}{2}\int_0^t \partial_z\psi_1(Z^\kappa_u,S_u,u)\mathrm{d}\langle Z^\kappa\rangle_u \\
= & \Psi(Z^\kappa_0,S_0,0) + \frac{1}{\kappa^2}\int_0^t \psi(Z^\kappa_u,S_u,u)\nu(S_u,u)^2\mathrm{d}u + M^\kappa_t
 \end{split}
\end{equation*}
for $t \in [0,T)$, where
\begin{equation*}
\begin{split}
 M^\kappa_t =& \frac{1}{\kappa}\int_0^t\psi_1(Z^\kappa_u,S_u,u)\mathrm{d}X^\alpha_u 
 +  \int_0^t \partial_s\Psi(Z^\kappa_u,S_u,u)\mathrm{d}S_u 
 \\ & +  \int_0^t \partial_t\Psi(Z^\kappa_u,S_u,u)\mathrm{d}u + \frac{1}{2} \int_0^t \partial_s^2 \Psi(Z^\kappa_u,S_u,u)\mathrm{d}\langle S \rangle_u \\ & + \frac{1}{\kappa} \int_0^t \partial_s\psi_1(Z^\kappa_u,S_u,u)\mathrm{d}\langle X^\alpha, S \rangle_u.
\end{split}
\end{equation*}
Here we have used (\ref{sko}), (\ref{marpsi}) and (\ref{genpsi}).
Since \begin{equation}\label{zbound}
 |Z^\kappa_t| \leq b(S_t,t),
\end{equation}
the condition (\ref{marg}) implies that
\begin{equation*}
 \lim_{t \to T}\Psi(Z^\kappa_t,S_t,t) = \Psi(Z^\kappa_0,S_0,0) +
\frac{1}{\kappa^2}\int_0^T \psi(Z^\kappa_u,S_u,u)\nu(S_u,u)^2\mathrm{d}u + M^\kappa_T.
\end{equation*}
Therefore,
\begin{equation*}
\begin{split}
&\sup_{t\in[0,T]}\left|\int_0^t \psi(Z^\kappa_u,S_u,u)\nu(S_u,u)^2\mathrm{d}u \right| \\
&\leq \kappa^2\sup_{t \in [0,T)}|\Psi(Z^\kappa_t,S_t,t)- \Psi(Z^\kappa_0,S_0,0)| + \kappa^2 \sup_{t \in [0,T]}|M^\kappa_t| \to 0
\end{split}
\end{equation*}
in probability as $\kappa \to 0$. \hfill//// \\

\begin{lem}
 \label{centered}
Let $\psi$ be a function satisfying the conditions of Lemma~\ref{centered0}.
Then  
\begin{equation*}
 \sup_{t \in [0,T]} \left|\int_0^t \psi(Z^\kappa_u,S_u,u)\mathrm{d}u \right| \to 0 
\end{equation*} 
in probability as $\kappa \to \infty$.
\end{lem}
{\it Proof: } We use Lemma~\ref{centered0} and (\ref{nondeg}).
For $n \geq 1$, let $\nu_n$ be a $C^{2,1}$ function on $(0,\infty)\times [0,T]$ such that
$\nu_n(s,t) = |\nu(s,t)|$ when $|\nu(s,t)| \geq 2/n$ and $|\nu_n(s,t)|\geq 1/n$ and
$|v(s,t)^2 - v_n(s,t)^2| \leq 4n^{-2}$ for all $(s,t)$.
Then,
\begin{equation*}
 \psi_n(x,s,t) := \frac{\psi(x,s,t)}{\nu_n(s,t)^2}
\end{equation*}
meets the conditions of Lemma~\ref{centered0} and so, 
\begin{equation*}
 \sup_{t \in [0,T]}\left|
\int_0^t \psi_n(Z^\kappa_u,S_u,u)\nu(S_u,u)^2\mathrm{d}u 
\right| \to 0
\end{equation*}
in probability as $\kappa \to 0$ for each $n$. Therefore,
\begin{equation*}
\begin{split}
  \sup_{t \in [0,T]}\left|
\int_0^t \psi(Z^\kappa_u,S_u,u)\mathrm{d}u 
\right| \leq & 
 \sup_{t \in [0,T]}\left|
\int_0^t \psi_n(Z^\kappa_u,S_u,u)\nu(S_u,u)^2\mathrm{d}u 
\right| \\ &+  4\left|
\int_0^T \psi(Z^\kappa_u,S_u,u)^2\mathrm{d}u 
\right|^{1/2} \int_0^T 1_{\{|\nu(S_u,u)| < 2n^{-1} \}}\mathrm{d}u,
\end{split}
\end{equation*}
which converges to 0 as $\kappa \to 0$ and then, $n \to \infty$ by (\ref{nondeg}) and (\ref{zbound}).
\hfill//// \\

\noindent
{\it Proof of Theorem~\ref{main}: }
Since a stable convergence is preserved under the localization and the Girsanov-Maruyama transformation, we can and do assume $\Theta = 0$
 without loss of generality.
Note that $h$ defined by (\ref{funs}) satisfies
\begin{equation}\label{heq}
 -\mathrm{sgn}(z) c(z,s,t) h(z,s,t) + \frac{1}{2}\partial_z h(z,s,t) = c(z,s,t)- \frac{1}{a(s,t)}
\end{equation} 
and
\begin{equation}\label{hb}
 h(b(s,t),s,t) = -1, \ \ h(-b(s,t),s,t) = 1.
\end{equation}
Let
\begin{equation*}
 H(x,s,t) = \lambda(s,t) \int_0^x h(z,s,t)\mathrm{d}z.
\end{equation*}
Then by the generalized It$\hat{\text{o}}$ formula, using (\ref{sko}), (\ref{heq}) and (\ref{hb}),
\begin{equation*}
 \begin{split}
 H(Z^\kappa_t, & S_t,t) -  H(Z^\kappa_0,S_0,0) \\
 =& \int_0^t \lambda(S_u,u)
h(Z^\kappa_u,S_u,u)\mathrm{d}Z^\kappa_u + \int_0^t \partial_s H(Z^\kappa_u,S_u,u)\mathrm{d}S_u \\ 
&+  \int_0^t \partial_tH(Z^\kappa_u,S_u,u)\mathrm{d}u + \frac{1}{2} \int_0^t \partial_s^2 H(Z^\kappa_u,S_u,u)\mathrm{d}\langle S \rangle_u \\ & + \int_0^t \partial_s (\lambda h)(Z^\kappa_u,S_u,u)\mathrm{d}\langle Z^\kappa, S \rangle_u + 
\frac{1}{2}\int_0^t \lambda(S_u,u)\partial_z h(Z^\kappa_u,S_u,u)\mathrm{d}\langle Z^\kappa\rangle_u \\
 =&   \frac{1}{\kappa}\int_0^t \lambda(S_u,u)h(Z^\kappa_u,S_u,u)\mathrm{d}X^\alpha_u + 
\int_0^t \lambda(S_u,u)\mathrm{d}[L^\kappa  + R^\kappa]_u \\
 & + \frac{1}{\kappa^2} \int_0^t \lambda(S_u,u) \left(c(Z^\kappa_u,S_u,u) - \frac{1}{a(S_u,u)}\right) \nu(S_u,u)^2 \mathrm{d}u  + N^\kappa_t,
 \end{split}
\end{equation*}
where
\begin{equation*}
\begin{split}
 N^\kappa_t = & \int_0^t \partial_s H(Z^\kappa_u,S_u,u)\mathrm{d}S_u +  \int_0^t \partial_tH(Z^\kappa_u,S_u,u)\mathrm{d}u \\
 &+ \frac{1}{2} \int_0^t \partial_s^2 H(Z^\kappa_u,S_u,u)\mathrm{d}\langle S \rangle_u + 
\frac{1}{\kappa}\int_0^t \partial_s (\lambda h)(Z^\kappa_u,S_u,u)\mathrm{d}\langle X^\alpha, S \rangle_u.
\end{split}
\end{equation*}
Therefore, 
\begin{equation*}
\begin{split}
 \kappa^{-1}&\left(\mathcal{E}^{b,c,\kappa}_t - \int_0^t \delta(S_u,u)\mathrm{d}\langle S \rangle_u \right) \\
= &
\kappa(H(Z^\kappa_t,S_t,t) - H(Z^\kappa_0,S_0,0)) - 
\int_0^t h(Z^\kappa_u,S_u,u)\varphi(S_u,u)\mathrm{d}u - \kappa N^\kappa_t \\ 
&+
\int_0^t (Z^\kappa_u - \lambda(S_u,u)\partial_s^2p^\alpha(S_u,u)h(Z^\kappa_u,S_u,u))\mathrm{d}S_u,
\end{split}
\end{equation*}
where $\varphi$ is a certain $C^{2,1}$ function.
The first term converges to $0$ uniformly on $[0,T]$ in probability due to (\ref{zbound}).
Since
\begin{equation*}
z \mapsto h(z,s,t), \ \ z \mapsto \partial_s(\lambda h)(z,s,t)
\end{equation*}
are odd functions and $z \mapsto g(z,s,t)$ is an even function, the second and third terms also converge to $0$ in probability by
Lemma~\ref{centered}.

It remains to show that
\begin{equation*}
 \int_0^\cdot (Z^\kappa_u-\lambda(S_u,u)\partial_s^2p^\alpha(S_u,u)h(Z^\kappa_u,S_u,u))\mathrm{d}S_u
\end{equation*}
converges stably to $W_Q$ in law on $C[0,T]$. By Theorem IX.7.3  of Jacod and Shiryaev~\cite{JS},
 it suffices to see that
\begin{equation*}
\begin{split}
& \int_0^t (Z^\kappa_u- \lambda(S_u,u)\partial_s^2p^\alpha(S_u,u)h(Z^\kappa_u,S_u,u))^2\sigma(S_u,u)^2\mathrm{d}u \to Q_t,\\
& \int_0^t (Z^\kappa_u- \lambda(S_u,u)\partial_s^2p^\alpha(S_u,u)h(Z^\kappa_u,S_u,u))\sigma(S_u,u)^2\mathrm{d}u \to 0
\end{split}
\end{equation*}
in probability for all $t \in [0,T]$, both of which follow from Lemma~\ref{centered}.
\hfill////
\\

\section{Optimal strategies}
\begin{figure}
 \includegraphics[width=6cm]{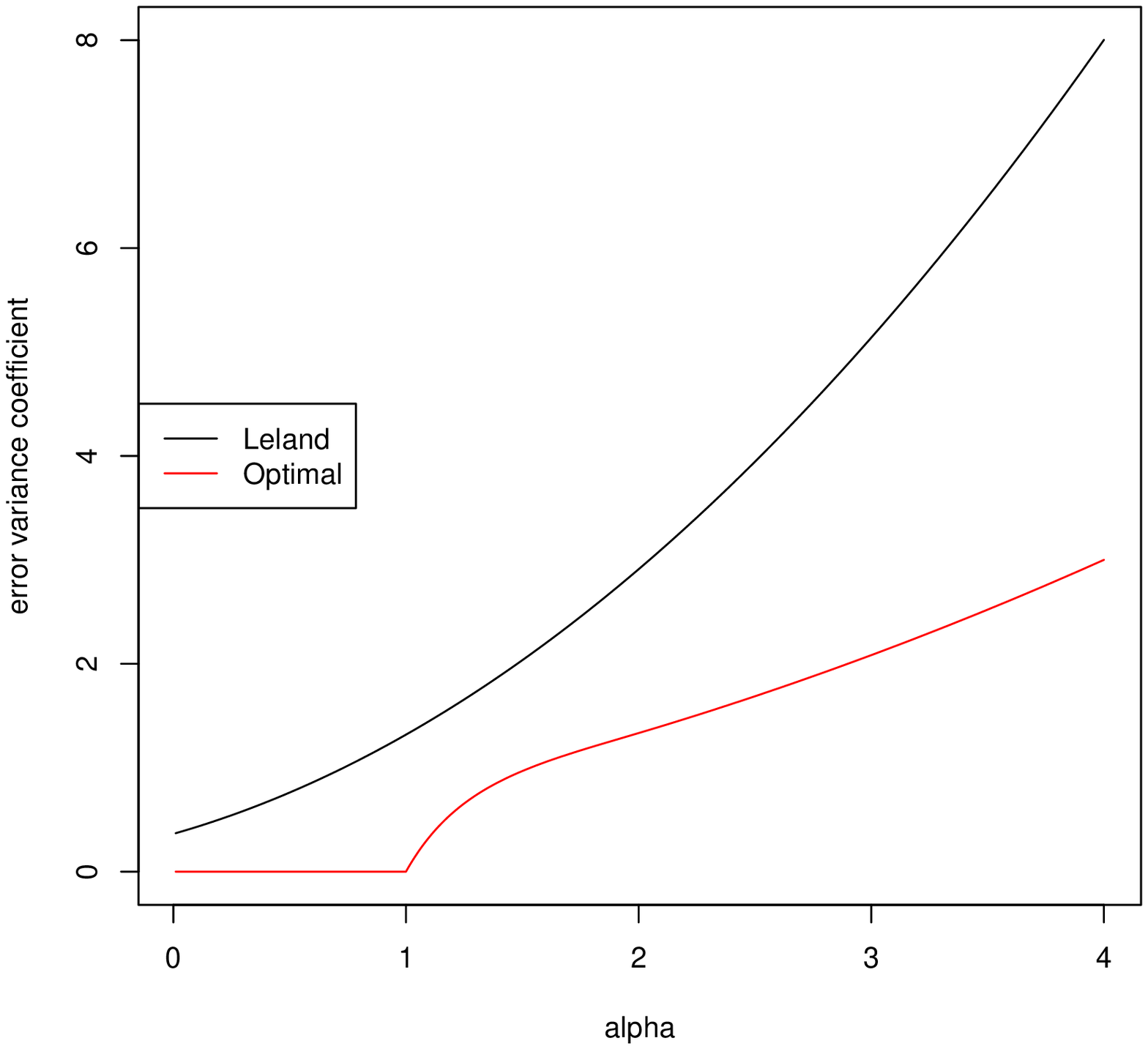}
 \includegraphics[width=6cm]{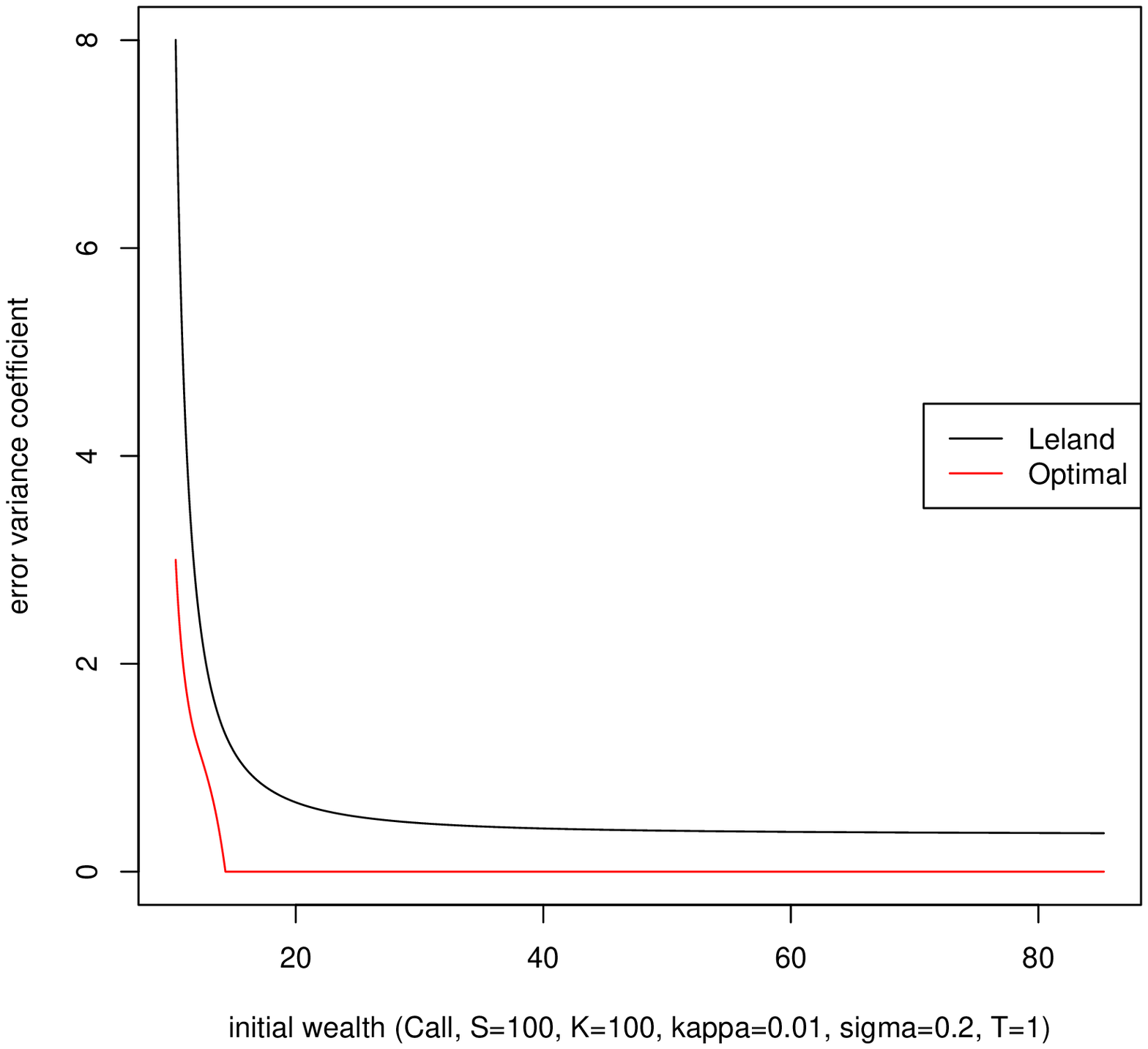}
\caption{Comparison between $\eta_L(\alpha)$ and $\eta_\dagger(\alpha)$.}\label{fig2}
\end{figure}

By the result of the previous section, the law of the hedging error associated with the strategy (\ref{gamma}) is approximated by the mixed normal distribution with mean
\begin{equation*}
 \int_0^T \delta(S_u,u)  \mathrm{d}\langle S \rangle_u
\end{equation*}
and variance
\begin{equation*}
 \kappa^2 \int_0^T \eta_{b,c}(S_u,u)  \mathrm{d}\langle S \rangle_u.
\end{equation*}
The function $\delta$ is determined by $\alpha$ and $a$ as (\ref{funs}). Therefore, in order to optimize our hedging strategy, we consider to minimize $\eta_{b,c}$ among $b$ and $c$ with $\alpha$ and $a$ fixed.
Let $a \in \mathcal{B}$  and
denote by $\mathcal{S}_a$ the set of $(b,c)$ with $b \in \mathcal{B}$, $c \in \mathcal{C}_b$ such that
\begin{equation}\label{aconst}
a(s,t) = 2 \int_0^{b(s,t)} g(x,s,t) \mathrm{d}x, \ \ g(x,s,t) = \exp\left\{ - 2 \int_0^{|x|}c(y,s,t) \mathrm{d}y \right\}.
\end{equation}
Let
\begin{equation*}
 \eta_\ast(s,t) = \inf_{(b,c) \in \mathcal{S}_a}\eta_{b,c}(s,t),
\end{equation*}
\begin{equation*}
 \gamma(s,t) = \lambda(s,t)\partial_s^2p^\alpha(s,t)
\end{equation*}
and
 \begin{equation*}
  \eta_\dagger(x) = \begin{cases}
\eta_2(x) & \text{ if } x \leq -2,\\ 
0 & \text{ if } -2 < x \leq 1,\\ 
\eta_1(x) & \text{ if } 1 < x < 2, \\
\eta_2(x) & \text{ if  } x \geq 2,
		   \end{cases}
 \end{equation*}
where
\begin{equation*}
 \eta_1(x) =  \frac{4}{3}\frac{(x + 2)^2(x-1)}{x^3(4-x)}, \ \ 
\eta_2(x) = \frac{1}{12}(x+2)^2.
\end{equation*}
\begin{thm} \label{main2}
For all $(s,t) \in (0,\infty) \times [0,T)$,
\begin{equation*}
\eta_\ast(s,t) = |\gamma(s,t)|^2\eta_\dagger \left( \frac{a(s,t)}{\gamma(s,t)}
\right)
\end{equation*}
if  $\gamma(s,t)\neq 0$, and 
\begin{equation*}
\eta_{\ast}(s,t)= \lim_{\epsilon \to 0} \epsilon^2 \eta_\dagger
\left(\frac{a(s,t)}{\epsilon}\right) = 
 \frac{a(s,t)^2}{12}
\end{equation*}
otherwise.
\end{thm}
To compare with Leland's strategy, assume $\gamma(s,t) > 0$ and $\lambda(s,t) = s$ and
take $ a(s,t) = \alpha \gamma(s,t)$ for a constant $\alpha$. 
Then $\delta = 0$ and  the limit variance is 
\begin{equation*}
\eta_\dagger(\alpha) \int_0^T |S_t\Gamma^\alpha_t|^2 \mathrm{d}\langle S \rangle_t. 
\end{equation*}
The minimized function $\eta_\dagger$ is continuous and nondecreasing on $[0,\infty)$ and less than half of $\eta_L$; see Figure~\ref{fig2}.
The aim of this section is to prove Theorem~\ref{main2}.
An explicit sequence $(b_n,c_n)$ which attains the infimum is given in the proof.
Fix for a while $(s,t) \in (0,\infty) \times [0,T]$ and for notational convenience, write
\begin{equation*}
\begin{split}
&a = a(s,t), \ \  b = b(s,t), \ \ c(x) = c(x,s,t), \ \ g(x) = g(x,s,t), \ \ h(x) = h(x,s,t),\\
& \gamma = \gamma(s,t), \ \ \eta_{b,c} = \eta_{b,c}(s,t).
\end{split}.
\end{equation*}
\begin{lem}
If $c$ is continuous, then $h^\prime$ is continuous and  for all $x \in (0,b)$,
\begin{equation} \label{ch}
h(x) > -1, \ \ h^\prime(x) \geq -\frac{2}{a}, \ \ c(x) = \frac{h^\prime(x) + 2/a}{2(1+h(x))}.
\end{equation}
\end{lem}
{\it Proof: }
Recall that
\begin{equation} \label{hmar}
  h(0) = 0, \ \ h(b) = -1.
\end{equation}
By (\ref{heq}), we have
\begin{equation}\label{heq1}
 2(1+h(x))c(x) = h^\prime(x) + \frac{2}{a}.
\end{equation}
Therefore $h^\prime$ is continuous on $[0,b]$. Further,
if there exists $ x \in (0,b)$ such that $h(x) \leq -1$, then $h^\prime(x) \leq -2/a < 0 $ since $c \geq 0$.
As a result, $h(\hat{x}) < -1$ for all $\hat{x} > x$. This contradicts (\ref{hmar}).
Thus we obtain $h(x) > -1$. From this and (\ref{heq1}) again, we conclude (\ref{ch}).
\hfill////
\begin{lem} \label{lem:xode}
If $c$ is continuous, then
\begin{equation*}
 \eta_{b,c} = 
 \frac{2}{a}\int_0^\infty(x(t)+\gamma(1 - x^\prime(t)))^2 e^{-2t/a}\mathrm{d}t,
\end{equation*}
where $x$ is the solution of the ordinary differential equation
\begin{equation} \label{xode}
 x^\prime = 1 + h(x), \ \ x(0) = 0.
\end{equation}
Further, the solution $x$ is $C^2$ and satisfies
\begin{equation} \label{xmar}
 x(\infty) = b, \ \ x^\prime(0) = 1, \ x^\prime(\infty) = 0, \ x^\prime > 0, \ \ x^{\prime \prime} \geq -\frac{2}{a}x^\prime.
\end{equation}
\end{lem}
{\it Proof: } By (\ref{ch}),
\begin{equation} \label{bg}
\int_0^b g(x)\mathrm{d}x = \int_0^b \frac{\mathrm{d}x}{1 + h(x)} \exp\left\{
-\frac{2}{a}\int_0^x \frac{\mathrm{d}y}{1 + h(y)}
\right\} = \int_0^{\hat{b}} e^{-2t/a} \mathrm{d}t,
\end{equation}
where
\begin{equation*}
 \hat{b} = \int_0^b \frac{\mathrm{d}x}{1+h(x)}.
\end{equation*}
Since (\ref{bg}) is equal to $a/2$ by definition, we conclude $\hat{b} = \infty$. As a result,
\begin{equation*}
\begin{split}
 \eta_{b,c} =& \frac{2}{a}\int_0^b \frac{(x-\gamma h(x))^2}{1+h(x)}\exp\left\{
-\frac{2}{a}\int_0^x \frac{\mathrm{d}y}{1 + h(y)} \right\}\mathrm{d}x \\
= & \frac{2}{a}\int_0^\infty(x(t) + \gamma (1- x^\prime(t)))^2 e^{-2t/a}\mathrm{d}t,
\end{split}
\end{equation*}
where $t \mapsto x(t)$ is the inverse function of
\begin{equation*}
 x \mapsto t(x) = \int_0^x\frac{\mathrm{d}y}{1+h(y)}.
\end{equation*}
The rest follows from (\ref{ch}) and (\ref{hmar}). \hfill////

\begin{lem}
If $c$ is continuous, then there corresponds an increasing convex $C^{2}$ function $y$ on $[0,1]$ such that
\begin{equation}\label{ymar}
 y(0) = 0, \ \ y^\prime(0) = \frac{a}{2}
\end{equation}
and
 \begin{equation*}
  \eta_{b,c} = \eta_a[y] := \int_0^1 \left(
 y(u) + \gamma + \frac{2\gamma}{a} (u-1)y^{\prime}(u)
\right)^2 \mathrm{d}u.
 \end{equation*}
\end{lem}
{\it Proof: }
Let $x$ be the solution of (\ref{xode}) and
\begin{equation*}
 y(u) = x\left(-\frac{a}{2}\log(1-u)\right) = x(v^{-1}(u)), \ \ 
 v(t) = 1- e^{-2t/a}.
\end{equation*}
Since $x(t) = y(v(t))$, we have
\begin{equation*}
 x^\prime(t) = \frac{2}{a} y^\prime(v(t))(1-v(t)), \ \ 
x^{\prime \prime}(t) = \frac{4}{a^2}y^{\prime\prime}(v(t))(1-v(t))^2 - \frac{2}{a} x^\prime(t).
\end{equation*}
From (\ref{xmar}), we obtain $y^{\prime \prime} \geq 0$ and $y^\prime > 0$.
Changing variable  $u = v(t)$, the result follows from Lemma~\ref{lem:xode}. \hfill////

\begin{lem}
Denote by $\mathcal{Y}_a$ the set of increasing convex functions $y$ on $[0,1]$ with (\ref{ymar}). 
Then
\begin{equation*}
 \inf_{y \in \mathcal{Y}_a}\eta_a[y] =  \lim_{x \to \gamma }x^2\eta_\dagger(a/x).
\end{equation*}
\end{lem}
{\it Proof: } It is easy to see that when $\gamma = 0$, the minimum of $\eta_a$ 
is attained by $y(u) = au/2$ and so, 
\begin{equation*}
 \inf_{y \in \mathcal{Y}_a}\eta_a[y] = \frac{a^2}{12}
=   \lim_{x \to \gamma }x^2\eta_\dagger(a/x).
\end{equation*}
Now, suppose $\gamma \neq 0$. Then
we have five cases: (1)~$-2 < a/\gamma < 1$, (2)~$a = \gamma$, (3)~$1 < a/\gamma < 2$, (4)~$a/\gamma \geq 2$ and (5)~$a/\gamma \leq -2$.\\

\noindent
Case 1). Assume $-2 < a/\gamma < 1$. For $\epsilon \in (0,1)$, define $y_\epsilon$ as
\begin{equation*}
 y_{\epsilon}(u) =
\begin{cases}
\gamma(1-u)^{-a/2\gamma} - \gamma & \text{ if } 0 \leq u \leq 1-\epsilon \\
 y_{\epsilon,0} + 
y_{\epsilon,0}^\prime(u-u_{\epsilon, 0}) & \text{ if } 1-\epsilon < u \leq 1, 
\end{cases}
\end{equation*}
where
\begin{equation*}
 u_{\epsilon,0} = 1-\epsilon, \ \ y_{\epsilon,0} = \gamma(1-u_{\epsilon,0})^{-a/2\gamma}-\gamma, \ \ 
y_{\epsilon,0}^\prime = \frac{a}{2}(1-u_{\epsilon,0})^{-a/2\gamma-1}. 
\end{equation*}
Then, $y_{\epsilon} \in \mathcal{Y}_a$ for all $\epsilon$.  Note that
\begin{equation*}
 y_{\epsilon}(u) +\gamma + \frac{2\gamma}{a}(u-1)y_\epsilon^\prime(u) = 
\gamma(1-u)^{-a/2\gamma} + \frac{2\gamma}{a}(u-1)\frac{a}{2}(1-u)^{-a/2\gamma-1} = 0
\end{equation*}
for $u \in (0,1-\epsilon)$. Therefore,
\begin{equation*}
\eta_a[y_{\epsilon}] = \eta(u_{\epsilon,0},y_{\epsilon,0}, y_{\epsilon,0}^\prime, 2\gamma/a),
\end{equation*}
where
\begin{equation*}
\eta(v,w,z,\beta) = \int_v^1 \left(w + z(u-v) + \gamma + \beta(u-1)z \right)^2 \mathrm{d}u.
\end{equation*}
By a straightforward calculation, 
\begin{equation}\label{etacal}
\begin{split}
 \eta(v,w,z,\beta) =  &
\frac{\beta^2- \beta + 1}{3}(1-v)^3
\left(z + \frac{3}{2}\frac{1-\beta}{\beta^2-\beta + 1} \frac{\gamma+w}{1-v}\right)^2
 \\ &+ \frac{(\beta + 1)^2}{4(\beta^2-\beta + 1)}(\gamma+w)^2(1-v).
\end{split}
\end{equation}
It follows that
\begin{equation}\label{eqn: eps to 0}
\eta(u_{\epsilon,0},y_{\epsilon,0}, y_{\epsilon,0}^\prime, 2\gamma/a) = O(\epsilon^{1-a/\gamma}) \to 0
\end{equation}
as $\epsilon \to 0$, which means that
\begin{equation*}
 \inf_{y \in \mathcal{Y}_a}\eta_a[y] = 0 = |\gamma|^2\eta_\dagger(a/\gamma).
\end{equation*}
\cai{Note that if $a/\gamma < -2$ then $y_\epsilon$ is not convex and if $a/\gamma \geq 1 $ then \eqref{eqn: eps to 0} doesn't hold. }
\\
\noindent
Case 2). Assume $a = \gamma$. For $\epsilon \in (0,1)$, define $y_{\epsilon}$ as
\begin{equation*}
 y_{\epsilon}(u) = 
\begin{cases}
\frac{1}{1-\epsilon} \left\{(1-u)^{-(1-\epsilon) /2} - 1\right\}  & \text{ if }  0 \leq u \leq 1-e^{-1/\epsilon^2} \\
 y_{\epsilon,0} + 
y_{\epsilon,0}^\prime(u-u_{\epsilon, 0}) & \text{ if } 1-e^{-1/\epsilon^2} < u \leq 1,
\end{cases}
\end{equation*}
where
\begin{equation*}
 u_{\epsilon,0} = 1-e^{-1/\epsilon^2}, \ \ y_{\epsilon,0} = \frac{1}{1-\epsilon}\left\{(1-u_{\epsilon,0})^{-(1-\epsilon)/2}-1\right\}, \ \ 
y_{\epsilon,0}^\prime = \frac{1}{2}(1-u_{\epsilon,0})^{-(1-\epsilon)/2-1}. 
\end{equation*}
Then, $y_{\epsilon} \in \mathcal{Y}_a$ for all $\epsilon$.  Note that
\begin{equation*}
 1 + y_{\epsilon}(u) + 2(u-1)y_{\epsilon}^\prime(u) = 
-\frac{\epsilon}{1-\epsilon}
+ \frac{\epsilon}{1-\epsilon}(1-u)^{-(1-\epsilon)/2}
\end{equation*}
for $u \in (0,1-\epsilon)$. Therefore,
\begin{equation*}
 \eta_a[y_\epsilon] = 
\frac{\epsilon^2}{(1-\epsilon)^2}
\int_0^{u_{\epsilon,0}}\left\{(1-u)^{-(1-\epsilon)/2}-1\right\}^2 \mathrm{d}u + 
\eta(u_{\epsilon,0},y_{\epsilon,0}, y_{\epsilon,0}^\prime, 2),
\end{equation*}
which converges to $0 = \eta_\dagger(1)$ as $\epsilon \to 0$ by (\ref{etacal}).\\

\noindent
Case 3). Assume $1 < a/\gamma < 2$.
For $ y \in \mathcal{Y}_a$, let
\begin{equation*}
 \varphi(u) = (1-u)^{a/2\gamma}(y(u) + \gamma).
\end{equation*}
Then, $\varphi(0) = \gamma$, $\varphi(1) = 0$ and
\begin{equation*}
 y + \gamma + \frac{2\gamma}{a}(u-1)y^\prime(u) = - \frac{2\gamma}{a}(1-u)^{1-a/2\gamma}\varphi^\prime(u).
\end{equation*}
By the Cauchy-Schwarz inequality, for $u_0 \in [0,1]$,
\begin{equation}\label{CS}
 \varphi(u_0)^2 =\left| \int_{u_0}^1\varphi^\prime(u)\mathrm{d}u\right|^2
\leq \int_{u_0}^1(1-u)^{2-a/\gamma}|\varphi^\prime(u)|^2\mathrm{d}u \int_{u_0}^1(1-u)^{a/\gamma-2}\mathrm{d}u.
\end{equation}
Therefore,
\begin{equation}\label{CS2}
\begin{split}
\int_{u_0}^1 \left(
 y + \gamma + \frac{2\gamma}{a}(u-1)y^\prime(u)
\right)^2 \mathrm{d}u & = \frac{4\gamma^2}{a^2}\int_{u_0}^1
(1-u)^{2-a/\gamma}|\varphi^\prime(u)|^2 \mathrm{d}u \\ &
\geq \frac{4\gamma^2}{a^2} \left(\frac{a}{\gamma}-1\right)(1-u_0)^{1-a/\gamma} \varphi(u_0)^2.
\end{split}
\end{equation}
Under the constraint $\varphi(1) = 0$, (\ref{CS2}) attains equality if and only if there exists $r \in \mathbb{R}$ such that
\begin{equation*}
 \varphi(u) = r(1-u)^{a/\gamma-1},
\end{equation*}
which corresponds to
\begin{equation} \label{opt}
 y(u) = r(1-u)^{a/2\gamma-1}-\gamma.
\end{equation}
For $\epsilon \in (0,1-u_0)$, define $y_{\epsilon}$ as
\begin{equation}\label{yep}
 y_\epsilon(u) = 
\begin{cases}
 au/2 & \text{ if } 0 \leq u \leq u_{0} \\
 r(1-u)^{a/2\gamma-1} - \gamma  & \text{ if } u_{0} < u \leq u_{\epsilon,0} \\
 y_{\epsilon,0} + 
y_{\epsilon,0}^\prime(u-u_{\epsilon, 0}) & \text{ if } u_{\epsilon,0} < u \leq 1,
\end{cases}
\end{equation} 
where $(u_0,r)$ is the solution of
\begin{equation}\label{u0r}
 \frac{a}{2} u_{0} = r(1-u_{0})^{a/2\gamma-1} -\gamma, \ \ 
\frac{a}{2} = r \left(1-\frac{a}{2\gamma}\right)(1-u_{0})^{a/2\gamma-2}
\end{equation}
and
\begin{equation*}
 u_{\epsilon,0} = 1-\epsilon, \ \ y_{\epsilon,0} = r(1-u_{\epsilon,0})^{a/2\gamma-1}-\gamma, \ \ 
y_{\epsilon,0}^\prime = r\left(1 - \frac{a}{2\gamma}\right)(1-u_{\epsilon,0})^{a/2\gamma-2}. 
\end{equation*}
The solution of (\ref{u0r}) uniquely exists and $u_0 \in (0,1)$ and $r> 0$; in fact
\begin{equation*}
 u_0 =\frac{4\gamma(a-\gamma)}{a(4\gamma-a)} > 0, \ \ 
1 - u_0 = \frac{4\gamma^2-a^2}{a(4\gamma-a)} > 0, \ \ 
r = \frac{a\gamma}{2\gamma-a}(1-u_0)^{2-a/2\gamma} > 0.
\end{equation*}
Then, $y_{\epsilon} \in \mathcal{Y}_a$ for all $\epsilon$. By a straightforward calculation,
\begin{equation*}
\begin{split}
 \eta_a[y_\epsilon] = &
\left(\gamma + \frac{a}{2}\right)^2\frac{u_0^3}{3} + 
\frac{4\gamma}{a^2}(a-\gamma)r^2\left((1-u_0)^{a/\gamma-1}-(1-u_{\epsilon,0})^{a/\gamma-1}\right)
\\ & + \eta(u_{\epsilon,0},y_{\epsilon,0},y_{\epsilon,0}^\prime,2\gamma/a)
\\  \to& 
\left(\gamma + \frac{a}{2}\right)^2\frac{u_0^3}{3} + 
\frac{4\gamma}{a^2}(a-\gamma)r^2(1-u_0)^{a/\gamma-1} \\
 = & \gamma^2 \eta_\dagger(a/\gamma)
\end{split}
\end{equation*}
 as $\epsilon \to 0$. Here we have used (\ref{etacal}) to observe $\eta(u_{\epsilon,0},y_{\epsilon,0},y_{\epsilon,0}^\prime,2\gamma/a) = O(\epsilon^{a/\gamma-1})$.
Now, let us show this is the infimum. Suppose there exists $y \in \mathcal{Y}_a$ such that $\eta_a[y] < \gamma^2\eta_\dagger(a/\gamma)$.
Since a convex function is approximated by piecewise linear convex functions arbitrarily close, we can and do assume $y$ itself is piecewise linear without loss of generality. 
Let 
\begin{equation*}
0 < u_1 < u_2 < \dots < u_n < 1
\end{equation*}
 be the discontinuity points of $y^\prime$. Let $u_0 = 0$ and $u_{n+1} = 1$. Denote
$y_i = y(u_i)$, $y^\prime_{i-} = \lim_{ u \uparrow u_i}y^\prime(u)$ and
$y^\prime_{i+} = \lim_{ u \downarrow u_i}y^\prime(u)$.
Note that 
\begin{equation*}
\frac{a}{2} = y^\prime_{0+} = y^\prime_{1-} < y^\prime_{1+} = y^\prime_{2-} < \dots  < y^\prime_{n+}
\end{equation*}
and $y^\prime_{i-}u_i \geq y_i$ for each $i$ by convexity.
Let $(v_i,r_i)$ be the solution of
\begin{equation*}
 y_i + y^\prime_{i+}(v_i-u_i) = r_i(1-v_i)^{a/2\gamma-1}-\gamma, \ \ 
y^\prime_{i+} = r_i\left(1-\frac{a}{2\gamma}\right)(1-v_i)^{a/2\gamma-2}.
\end{equation*}
The solution uniquely exists and $v_i \in (0,1)$ and $r_i > 0$; in fact
\begin{equation}\label{vi}
\begin{split}
& v_i = \frac{\gamma(2y^\prime_{i+}+a-2\gamma) + (2\gamma-a)(y^\prime_{i+}u_i -y_i)}{y^\prime_{i+}(4\gamma-a)} > 0, \\
& 1-v_i = \frac{(2\gamma-a)(\gamma +y_i + (1-u_i)y^\prime_{i+})}{y^\prime_{i+}(4\gamma-a)} > 0.
\end{split}
\end{equation}
Further, it holds $v_i < v_{i+1}$ because
\begin{equation*}
\begin{split}
  v_{i+1}-v_i =& 
\frac{2\gamma+(2\gamma-a)u_{i+1}}{4\gamma-a}-\frac{(2\gamma-a)(\gamma+y_{i+1})}{y^\prime_{i+1+}(4\gamma-a)}
\\ & -\frac{2\gamma+(2\gamma-a)u_i}{4\gamma-a}+\frac{(2\gamma-a)(\gamma+y_i)}{y^\prime_{i+}(4\gamma-a)} \\
> &
\frac{2\gamma+(2\gamma-a)u_{i+1}}{4\gamma-a}-\frac{(2\gamma-a)(\gamma+y_{i+1})}{y^\prime_{i+}(4\gamma-a)} \\
& -\frac{2\gamma+(2\gamma-a)u_i}{4\gamma-a}+\frac{(2\gamma-a)(\gamma+y_i)}{y^\prime_{i+}(4\gamma-a)} \\
=&
\frac{(2\gamma-a)(u_{i+1}-u_i)}{4\gamma-a} - \frac{(2\gamma-a)(y_{i+1}-y_i)}{y^\prime_{i+}(4\gamma-a)} = 0.
\end{split}
\end{equation*}
In particular if $v_{i+1} \leq u_{i+1}$, then $v_i < u_{i+1}$. 
This implies that the set
\begin{equation*}
\mathcal{I} := \{i \in \{0,1,\dots,n\} ; u_i \leq v_i < u_{i+1}\}
\end{equation*}
is not empty.
 In fact, if $v_n < u_n$, then $v_{n-1} < u_n$. If $v_{n-1} < u_{u-1}$, then
$v_{n-2} < u_{u-1}$. If $\mathcal{I}$ is empty, then we conclude $v_0 < u_0 = 0$ by induction, which contradicts (\ref{vi}).
Now, let
\begin{equation*}
 k = \min\{i \in \{0,1,2,\dots,n\} ; u_i \leq v_i < u_{i+1}\}.
\end{equation*}
Then $v_{i-1} \geq u_{i}$ for all $i \leq k$.
For $z \geq y^\prime_{i-}$, define $y_i(\cdot,z)$ by
\begin{equation}\label{yz}
 y_i(u,z) = \begin{cases}
	   y(u) & \text{ if } 0 \leq u \leq u_i \\
y_i + z(u-u_i) & \text{ if } u_i < u \leq v_i(z) \\
r_i(z)(1-u)^{a/2\gamma-1}-\gamma & \text{ if } v_i(z) < u \leq 1,
	  \end{cases}
\end{equation}
where
$(v_i(z),r_i(z))$ is the solution of
\begin{equation*}
 y_i + z(v_i(z)-u_i) = r_i(z)(1-v_i(z))^{a/2\gamma-1}-\gamma, \ \ 
z = r_i(z)\left(1-\frac{a}{2\gamma}\right)(1-v_i(z))^{a/2\gamma-2}.
\end{equation*}
The solution uniquely exists as before and we have
\begin{equation*}
 v_i(y^\prime_{i+}) = v_i, \ \ r_i(y^\prime_{i+}) = r_i
\end{equation*}
 and
\begin{equation*}
 v_i(y^\prime_{i-}) = v_{i-1}, \ \ r_i(y^\prime_{i-}) = r_{i-1}.
\end{equation*}
Since
\begin{equation*}
v_i(z) =
 \frac{2\gamma+(2\gamma-a)u_i}{4\gamma-a}-\frac{(2\gamma-a)(\gamma+y_i)}{z(4\gamma-a)},
\end{equation*}
$v_i(z)$ is an increasing function. This implies that
$v_i(z) \geq v_{i-1} \geq u_i$ for all $i \leq k$.
Note also that  $y_k(u,y^\prime_{k+}) = y(u)$ for $u \in [0,v_k]$.
Recall that a function of the form (\ref{opt})
for $u \geq v_k$  minimizes (\ref{CS2}) with $u_0 = v_k$. This implies that
\begin{equation*}
 \eta_a[y_k(\cdot,y^\prime_{k+})] < \eta_a[y].
\end{equation*}
Further, 
\begin{equation}\label{numer}
\begin{split}
& \partial_z \left\{\eta_a[y_i(\cdot,z)] \right\} \\
& = \frac{(4\gamma z(1-u_i) + (2\gamma-a)(\gamma+y_i))(2\gamma z(1-u_i) - (2\gamma-a)(\gamma+y_i))^2}{3z^2a^2(4\gamma-a)}\\
& >  0 
\end{split}
\end{equation}
for all  $z \geq y^{\prime}_{i-}$ (see Appendix).
It means that
\begin{equation*}
 \eta_a[y_k(\cdot,y^\prime_{k-})] <  \eta_a[y_k(\cdot,y^\prime_{k+})] < \eta_a[y].
\end{equation*}
The function $y_k(\cdot,y^{\prime}_{k-})$ is continuously differentiable at $u_k$ and coincides with $y_{k-1}(\cdot,y^{\prime}_{k-1+})$.
Again by (\ref{numer}), we have
\begin{equation*}
 \eta_a[y_{k-1}(\cdot,y^\prime_{k-1-})] <   \eta_a[y_{k-1}(\cdot,y^\prime_{k-1+})] < \eta_a[y].
\end{equation*}
We can repeat this argument to conclude
\begin{equation*}
  \eta_a[y_1(\cdot,y^\prime_{1-})]  < \eta_a[y].
\end{equation*}
Note that $y^\prime_{1-} = a/2$ and so, $y_1(\cdot,y^\prime_{1-})$ coincides with the limit of $y_\epsilon$ defined by (\ref{yep}) as $\epsilon \to 0$.
It is not difficult see $  \eta_a[y_1(\cdot,y^\prime_{1-})] = \gamma^2\eta_\dagger(a/\gamma)$, which contradicts how $y$ was chosen. This completes Case 3.\\

\noindent
Case 4). Assume $a/\gamma \geq 2$. Let $\hat{y}(u) = au/2$. Then,
 $\hat{y} \in \mathcal{Y}_a$ and $\eta_a[\hat{y}] = \gamma^2 \eta_\dagger(a/\gamma)$.
Suppose there exists $y \in \mathcal{Y}_a$ such that  $\eta_a[y] < \gamma^2 \eta_\dagger(a/\gamma)$.
Since a convex function is approximated by piecewise linear convex functions arbitrarily close, we can and do assume $y$ itself is piecewise linear without loss of generality. 
Let  $u_0 \in (0,1)$ be the last point where $y^\prime$ jumps. Denote $y_0 = y(u_0)$, 
$y^\prime_{-} = \lim_{ u \uparrow u_0}y^\prime(u)$ and
$y^\prime_{+} = \lim_{ u \downarrow u_0}y^\prime(u)$.
For $z \geq y^\prime_-$, define $y(\cdot,z)$ as
\begin{equation*}
 y(u,z) = \begin{cases}
y(u) & \text{ if } 0 \leq u \leq u_0 \\
y_0 + z(u-u_0) & \text{ if } u_0 < u \leq 1.	   
	  \end{cases}
\end{equation*}
Note that $y(\cdot, y^\prime_+) = y$. As seen before,
\begin{equation*}
 \int_{u_0}^{1}
\left(y(u,z) + \gamma + \frac{2\gamma}{a} (u-1)y^\prime(u,z)\right)^2
\mathrm{d}u = \eta(u_0,y_0,z,2\gamma/a)
\end{equation*}
and  $\eta$ is given by (\ref{etacal}). Since $1 - \beta = 1- 2\gamma/a  \geq 0$, the first term of (\ref{etacal}) is minimized by
$z = y^\prime_-$ on the region $z \geq y^\prime_-$. This implies
\begin{equation*}
 \eta_a[y(\cdot,y^\prime_-)] < 
 \eta_a[y(\cdot,y^\prime_+)] = \eta_a[y].
\end{equation*}
The function $y(\cdot,y^\prime_-)$ is continuously differentiable at $u_0$.
We can repeat the same argument with $y$ replaced by $y(\cdot,y^\prime_-)$ to obtain a smaller value of $\eta_a$.
Eventually, all discontinuity points are removed and the final product coincides with $\hat{y}(u) = au/2$. This contradicts how 
$y$ was chosen. This completes Case~4. \\

\noindent
Case 5). Assume $a/\gamma \leq -2$.
The idea is the same as in the previous case.
Let $\hat{y}(u) = au/2$. As before, $\hat{y} \in \mathcal{Y}_a$ and $\eta_a[\hat{y}] = \gamma^2 \eta_\dagger(a/\gamma)$.
Suppose there exists $y \in \mathcal{Y}_a$ such that  $\eta_a[y] < \gamma^2 \eta_\dagger(a/\gamma)$.
Since a convex function is approximated by piecewise linear convex functions arbitrarily close, we can and do assume $y$ itself is piecewise linear without loss of generality. 
Let  $u_0 \in (0,1)$ be the last point where $y^\prime$ jumps. Denote $y_0 = y(u_0)$, 
$y^\prime_{-} = \lim_{ u \uparrow u_0}y^\prime(u)$ and
$y^\prime_{+} = \lim_{ u \downarrow u_0}y^\prime(u)$.
For $z \geq y^\prime_-$, define $y(\cdot,z)$ as
\begin{equation*}
 y(u,z) = \begin{cases}
y(u) & \text{ if } 0 \leq u \leq u_0 \\
y_0 + z(u-u_0) & \text{ if } u_0 < u \leq 1.	   
	  \end{cases}
\end{equation*}
 As seen before, $y(\cdot, y^\prime_+) = y$ and
\begin{equation*}
 \int_{u_0}^{1}
\left(y(u,z) + \gamma + \frac{2\gamma}{a} (u-1)y^\prime(u,z)\right)^2
\mathrm{d}u = \eta(u_0,y_0,z,2\gamma/a)
\end{equation*}
with  $\eta$ given by (\ref{etacal}). 
Note that $y^\prime_- \geq a/2$ by the convexity \cai{of} $y$. Therefore,
the same argument as in the previous case remains valid here once we prove that
\begin{equation}\label{quadratic}
\cai{-\frac{3}{2}\frac{1-\beta}{\beta^2 - \beta + 1} \frac{\gamma + w}{1-v} 
\leq \frac{a}{2}}
\end{equation}
for $\beta=2\gamma/a$, $v = u_0$ and $w = y_0$.
By the convexity, we have $y_0 \geq u_0a/2$. Note also that
\begin{equation*}
\partial_v \left\{
\frac{\gamma + va/2}{1-v} \right\}= \frac{a+2\gamma}{2(1-v)^2} \geq 0.
\end{equation*}
Therefore,
\begin{equation*}
\begin{split}
 -\frac{3}{2}\frac{1-\beta}{\beta^2 - \beta + 1} \frac{\gamma + y_0}{1-u_0} 
&\leq -\frac{3}{2}\frac{1-\beta}{\beta^2 - \beta + 1} \frac{\gamma + u_0a/2}{1-u_0} \\
&\leq -\frac{3}{2}\frac{1-\beta}{\beta^2 - \beta + 1} \gamma.
\end{split}
\end{equation*}
To show (\ref{quadratic}), it suffices then to see
\begin{equation*}
 \frac{a}{2} + \frac{3}{2}\frac{1-\beta}{\beta^2 - \beta + 1} \gamma
= \frac{a(2-\beta)(1 + \beta)}{4(\beta^2 - \beta + 1)}\geq 0.
\end{equation*}
Now, we can deduce a contradiction as in the previous case, which completes the proof.
 \hfill////
\\

\noindent
{\it Proof of Theorem~\ref{main2}: }
In order to bound the infimum of $\eta_{b,c}$, it suffices to consider $c$ which is continuous. 
Then by the preceding lemmas, we conclude
\begin{equation*}
 \eta_{b,c}(s,t) \geq \lim_{x \to \gamma(s,t)} x^2\eta_\dagger(a(s,t)/x)
\end{equation*}
for any $(b,c) \in \mathcal{S}_a$. It remains to show that  the lower bound is asymptotically attained by a sequence $(b_n,c_n) \in \mathcal{S}_a$.
Let
\begin{equation*}
\begin{split}
 &l(s,t) =  \lim_{x \to \gamma(s,t)}\frac{2x^2(a(s,t)/x-1)_+}{4x-a(s,t)}, \\ 
&r_n(s,t) = l(s,t)+(2\gamma(s,t)-a(s,t))_+ e^{n^2} + |\gamma(s,t)|\left\{
\left(1 - \frac{1}{n}\right) \wedge (2\gamma(s,t) + a(s,t))_-
\right\}^{1/n}.
\end{split}
\end{equation*}
Let $\varphi_n$ be a smooth function such that $\varphi_n(x) = 1$ if  $|x-1|\geq 1/n$ and $\varphi_n(1) = 1- 1/n$.
Denote 
\begin{equation*}
 \psi_n(s,t) = \lim_{x \to \gamma(s,t)}\varphi_n(a(s,t)/x).
\end{equation*}
Define $b_n$ and $c_n$ as
\begin{equation}\label{optimal}
 \begin{split}
  &b_n(s,t) = r_n(s,t) + \frac{\gamma(s,t) + \psi_n(s,t)r_n(s,t)}{\gamma(s,t) + \psi_n(s,t)l(s,t)} \left(
\frac{a(s,t)}{2} - l(s,t)
\right)\\
& c_n(x,s,t) = \frac{a(s,t)\psi_n(s,t)+2\gamma(s,t)}{2(a(s,t)-2l(s,t))}
\frac{1}{\gamma(s,t)+\psi_n(s,t)|x|}
1_{[l(s,t), r_n(s,t))}(|x|).
 \end{split}
\end{equation}
Note that 
\begin{enumerate}
\item if $\gamma(s,t) = 0$, then $l(s,t) = r_n(s,t) = 0$ and so, 
\begin{equation*}
b_n(s,t) = \frac{a(s,t)}{2}, \ \ c_n(x,s,t) = 0,
\end{equation*}
\item if $\gamma(s,t) \neq 0$ and $a(s,t)/\gamma(s,t) \leq -2$, then $r_n(s,t) = l(s,t)$ again and so,
\begin{equation*}
b_n(s,t) = \frac{a(s,t)}{2}, \ \ c_n(x,s,t) = 0,
\end{equation*} 
 \item  if $\gamma(s,t) \neq 0$ and  $-2 < a(s,t)/\gamma(s,t) < 1$, then $l(s,t) = 0$ and for sufficiently large $n$,
\begin{equation}\label{degen0}
\begin{split}
& b_n(s,t) = r_n(s,t) + \frac{a(s,t)}{2\gamma(s,t)}(\gamma(s,t) + r_n(s,t)),\\
& c_n(x,s,t) = \frac{a(s,t)+2\gamma(s,t)}{2a(s,t)} \frac{1}{\gamma(s,t)+|x|} 1_{\{|x| < r_n(s,t)\}},
\end{split}
\end{equation}
\item 
if $a(s,t) = \gamma(s,t)$, then $l(s,t) = 0$ and
\begin{equation}\label{degen1}
\begin{split}
& b_n(s,t) = r_n(s,t) + \frac{1}{2}\left(\gamma(s,t) + \left(1-\frac{1}{n}\right)r_n(s,t)\right),\\ 
& c_n(x,s,t) = \frac{3n-1}{2(n\gamma(s,t) + (n-1)|x|)} 1_{\{|x| < r_n(s,t)\}},
\end{split}
\end{equation}
\item
if $\gamma(s,t) \neq 0$ and $1 < a(s,t)/\gamma(s,t) < 2$, then for sufficiently large $n$,
\begin{equation*}
\begin{split}
& b_n(s,t) = r_n(s,t) + \frac{(2\gamma(s,t)-a(s,t))(2\gamma(s,t)+a(s,t))}{2(4\gamma(s,t)-a(s,t))}\frac{\gamma(s,t)+r_n(s,t)}{\gamma(s,t)+l(s,t)},\\
& c_n(x,s,t) = \frac{4\gamma(s,t)-a(s,t)}{2(2\gamma(s,t)-a(s,t))} \frac{1}{\gamma(s,t)+|x|} 1_{\{l(s,t) \leq |x| < r_n(s,t)\}},
\end{split}
\end{equation*}
and 
\item if $a(s,t)/\gamma(s,t) \geq 2$, then $r_n(s,t) = l(s,t)$ and
\begin{equation*}
 b_n(s,t) = \frac{a(s,t)}{2},  \ \ c_n(x,s,t) = 0.
\end{equation*}
\end{enumerate}
It is straightforward to see
\begin{equation*}
\int_0^{b_n(s,t)}g_n(x,s,t)\mathrm{d}x = \frac{a(s,t)}{2}, \ \ 
g_n(x,s,t) = \exp\left\{
-2 \int_0^x c_n(y,s,t)\mathrm{d}y
\right\}.
\end{equation*}
Although $b_n$ is not necessarily $C^{2,1}$ and
 $c_n$ does not satisfy the one-sided Lipschitz condition due to the discontinuity at $r_n(s,t)$, 
they can be approximated arbitrarily close by smooth functions.
Let $h_n$ be the associated $h$ function with $b_n$ and $c_n$.

If $-2 < a(s,t)/\gamma(s,t) < 1$, then for sufficiently large $n$,
\begin{equation*}
 g_n(x,s,t) = 
\left(\frac{\gamma(s,t) + x\wedge r_n(s,t)}{\gamma(s,t)}\right)^{-1-2\gamma(s,t)/a(s,t)} 
\end{equation*}
and by a straightforward calculation,
\begin{equation*}
 h_n(x,s,t) = \begin{cases}
x/\gamma(s,t) & \text{ if } 0 \leq x \leq r_n(s,t), \\
r_n(s,t)/\gamma(s,t) -2(x-r_n(s,t))/a(s,t) & \text{ if } x > r_n(s,t).	       
	      \end{cases}
\end{equation*}
Therefore,
\begin{equation*}
\begin{split}
 \eta_{b_n,c_n}(s,t) = & \frac{2}{a(s,t)}
\left(1 + \frac{2\gamma(s,t)}{a(s,t)}\right)^2
\frac{(b_n(s,t)-r_n(s,t))^3}{3}\left(1+ \frac{r_n(s,t)}{\gamma(s,t)}\right)^{-1-2\gamma(s,t)/a(s,t)} \\ = & 
O((\gamma(s,t)+r_n(s,t))^{2-2\gamma(s,t)/a(s,t)})  \to 0 \\ = & \gamma(s,t)^2\eta_\dagger(a(s,t)/\gamma(s,t))
\end{split}
\end{equation*}
as $n\to \infty$.

If $a(s,t) = \gamma(s,t)$, then
\begin{equation*}
 g_n(s,t) = \left(
1 + \left(1-\frac{1}{n}\right) \frac{x \wedge r_n(s,t)}{\gamma(s,t)}
\right)^{-(3n-1)/(n-1)}.
\end{equation*}
Again by a straightforward calculation, we find that
\begin{equation*}
 h_n(x,s,t) = \begin{cases}
\left(1-\frac{1}{n}\right) \frac{x}{\gamma(s,t)} & \text{ if } 0 \leq x \leq r_n(s,t), \\
\left(1-\frac{1}{n}\right) \frac{r_n(s,t)}{\gamma(s,t)} - \frac{2}{a(s,t)}(x-r_n(s,t)) & \text{ if } x > r_n(s,t)
	      \end{cases}
\end{equation*}
and
\begin{equation*}
 \eta_{b_n,c_n}(s,t) \to 0 = \gamma(s,t)^2\eta_\dagger(a(s,t)/\gamma(s,t)).
\end{equation*}
as $n \to \infty$.

If $1 < a(s,t)/\gamma(s,t) < 2$, then for sufficiently large $n$,
\begin{equation*}
 g_n(x,s,t) = \left(\frac{\gamma(s,t) + x \wedge r_n(s,t)}
{\gamma(s,t) + x \wedge l(s,t)}\right)^{-(4\gamma(s,t)-a(s,t))/(2\gamma(s,t)-a(s,t))}.
\end{equation*}
In this case,
\begin{equation*}
 h_n(x,s,t) = \begin{cases}
-\frac{2}{a(s,t)}x  & \text{if }  0 \leq x \leq l(s,t), \\ 
\left(\frac{2}{a(s,t)}-\frac{1}{\gamma(s,t)}\right)x + \frac{2\gamma(s,t)}{a(s,t)}-2	& \text{if }  l(s,t) < x \leq r_n(s,t),  \\
\left(\frac{2}{a(s,t)}-\frac{1}{\gamma(s,t)}\right)r_n + \frac{2\gamma(s,t)}{a(s,t)}-2	- \frac{2}{a(s,t)}(x-r_n(s,t))& \text{if } x >  r_n(s,t).       
	      \end{cases}
\end{equation*}
This also 
can be shown by a direct computation; an easier method is however to see this $h_n$ recovers $c_n$ by (\ref{ch}).
Further, this function is obtained by (\ref{xode}) and changing variable $u = 1-\exp\{-2t/a\}$
from $y_\epsilon$ defined by (\ref{yep}) with a certain $\epsilon$ depending on $n$. Consequently, 
\begin{equation*}
 \eta_{b_n,c_n}(s,t) \to \gamma(s,t)^2\eta_\dagger(a(s,t)/\gamma(s,t))
\end{equation*}
as $n\to \infty$.

Finally if $a(s,t)\geq 2|\gamma(s,t)|$ or $\gamma(s,t) = 0$, 
then it is easy to see
\begin{equation*}
 g_n(x,s,t) = 1,  \ \ h_n(x,s,t) = - \frac{2}{a(s,t)}x
\end{equation*}
and
\begin{equation*}
 \eta_{b_n,c_n}(s,t) = \lim_{x \to \gamma(s,t)}x^2\eta_\dagger(a(s,t)/x).
\end{equation*}
This completes the proof. \hfill////\\

\section{Concluding remarks}
We have constructed hedging strategies for a general European option under a general local volatility model with small but nonnegligible transaction costs. The strategies are based on Leland's idea of modifying volatility to absorb the transaction costs. We have proved the stable convergence of the associated hedging error process
to a conditionally Gaussian semimartingale.
Further, we have shown that the infimum of the asymptotic variance has an explicit form and is asymptotically attained by a sequence of explicit strategies.
Here some remarks on the results are given in order.\\

\noindent
-- The law of the hedging error associated with the asymptotically optimal strategy (\ref{optimal}) is approximated by a mixed normal distribution with mean
\begin{equation*}
 \int_0^T \left\{
\frac{|\gamma(S_t,t)|}{a(S_t,t)} - \frac{1}{\alpha(S_t,t)}
\right\}
|\partial_s^2p^\alpha(S_t,t)|\mathrm{d}\langle S \rangle_t
\end{equation*}
and variance
\begin{equation*}
\kappa^2 \int_0^T \gamma(S_t,t)^2\eta_\dagger\left(
\frac{a(S_t,t)}{\gamma(S_t,t)}\right) \mathrm{d}\langle S \rangle_t.
\end{equation*}
For a given $\alpha$, we can do a further optimization with respect to $a$ under, say, a mean-variance criterion.
For a constant $A>0$, the optimized function $a^\ast$ is given by
\begin{equation*}
a^\ast =  \mathrm{argmin}_{a>0}\left\{
\frac{|\gamma|}{a} + A \lambda |\gamma|\eta_\dagger\left(
\frac{a}{\gamma}
\right)
\right\}
\end{equation*} 
when $\gamma \neq 0$.
Since $\eta_\dagger(a/\gamma) = 0$ for $-2 \leq a/\gamma \leq 1$, 
we have $a^\ast > \max\{\gamma, -2\gamma\}$.\\

\noindent
-- Now, we can consider
 optimization with respect to $\alpha$ under a constraint on the initial capital  $p^\alpha(S_0,0)$.
Finding an efficient algorithm remains for future research.
Since $\gamma = \lambda \partial_s^2 p^\alpha$,
the solution would involve such a nonlinear \cai{PDE} that is given in Barles and Soner~\cite{BS}, where a scaling limit of exponential utility indifference price is considered.
For mixed normal distributions, the mean-variance criterion above will give the same result as the exponential utility maximization. However the nonlinear PDE will not be exactly the same as the one in Barles and Soner~\cite{BS} because the ways how to scale transaction cost coefficient and risk aversion parameter are different. In fact, we have a regular control part in the asymptotically optimal strategy, while it is purely singular in Barles and Soner~\cite{BS}.
The difference from the asymptotic analysis in Whalley and Wilmott~\cite{WW} and
Soner and Touzi~\cite{ST} also lies on how to scale the transaction cost coefficient. \\

\noindent
-- Another stream of research on Leland's strategy is to fix $\kappa$ and let $\alpha \to 0$.
The initial capital required by the strategy typically converges to the super replication price as $\alpha \to 0$.
Therefore it is natural to expect that the strategy asymptotically super-replicates the payoff.
The original method of Leland however fails as shown by Kabanov and Safarian~\cite{KS}.
It is not difficult to get an intuition on this; the hedging error variance of Leland's strategy is approximated by
\begin{equation*}
 \kappa^2 \eta_L(\alpha)\int_0^T |S_t\Gamma^\alpha_t|^2\mathrm{d}\langle S \rangle_t
\end{equation*}
when $\kappa$ is small.
This does not converges to $0$ as $\alpha \to 0$ with $\kappa$ fixed because $\eta_L(0) = 1-2/\pi > 0$.
On the other hand, since $\eta_\dagger(0) = 0$, the asymptotically optimal strategy (\ref{optimal}) with $a = \alpha|\gamma|$ works also in this asymptotic framework. More specifically, we can show that the trading strategy
\begin{equation*}
 \mathrm{d}X_t = \mathrm{sgn}(X^\alpha_t-X_t)\frac{\alpha + 2}{2\alpha}
\frac{\nu(S_t,t)^2}{\kappa \gamma(S_t,t) + |X^\alpha_t - X_t|} \mathrm{d}t
\end{equation*}  
asymptotically replicates the payoff as $\alpha \to 0$ with $\kappa > 0$ fixed, 
when $f(s) = s\log(s)$ and $\lambda(s,t) = s$ that make 
$\nu$ and $\gamma$ constant.
To treat a more general payoff and to
show its optimality remain for future research.\\

\noindent
{\bf Acknowledgement: }
\cai{J. Cai thanks Professors Mathieu Rosenbaum and Peter Tankov for making this collaboration possible.} M. Fukasawa is grateful to Professors Chiaki Hara, Toshiki Honda, Masaaki Kijima, Shigeo Kusuoka and Jun Sekine for their helpful comments and suggestions.
This work is supported by Japan Society for the Promotion of Science, KAKENHI Grant Numbers
24684006 and 25245046.

\newpage
\begin{appendices}
\section{Note on computation for (\ref{numer})}

The computation for (\ref{numer}) is straightforward but lengthy.
Here we present a Maple worksheet to check it.

\begin{verbatim}
> v:=z ->2*g /(4*g-a) +  (2*g-a)*(u -  (g+y)/z) /(4*g-a);
                         /    g + y\
               (2 g - a) |u - -----|
       2 g               \      z  /
z -> ------- + ---------------------
     4 g - a          4 g - a       

> eta :=  16*g^3*(a-g)*z^2*(1-v(z))^3/(a^2*(2*g-a)^2)
+(1/3)*(1+2*g/a)^2*z^2*(v(z)^3-u^3)
+(1+2*g/a)*z*(g+y-z*(u+2*g/a))*(v(z)^2-u^2)
+(v(z)-u)*(g+y-z*(u+2*g/a))^2;
               3          2            3
           16 g  (a - g) z  (1 - v(z)) 
      z -> -----------------------------
                    2          2        
                   a  (2 g - a)         

                      2                 
           1 /    2 g\   2 /     3    3\
         + - |1 + ---|  z  \v(z)  - u /
           3 \     a /                  

           /    2 g\   /          /    2 g\\ /     2    2\
         + |1 + ---| z |g + y - z |u + ---|| \v(z)  - u /
           \     a /   \          \     a //              

                                            2
                       /          /    2 g\\ 
         + (v(z) - u) |g + y - z |u + ---|| 
                       \          \     a // 

> latex(diff(eta(z), z));
\end{verbatim}
\begin{equation*}
\begin{split}
&32\,{g}^{3} \left( a-g \right) z \left( 1-2\,{\frac {g}{4\,g-a}}-
 \left( 2\,g-a \right)  \left( u-{\frac {g+y}{z}} \right)  \left( 4\,g
-a \right) ^{-1} \right) ^{3}{a}^{-2} \left( 2\,g-a \right) ^{-2} \\ &-48\,
{g}^{3} \left( a-g \right)  \left( 1-2\,{\frac {g}{4\,g-a}}- \left( 2
\,g-a \right)  \left( u-{\frac {g+y}{z}} \right)  \left( 4\,g-a
 \right) ^{-1} \right) ^{2} \\ & \left( g+y \right) {a}^{-2} \left( 2\,g-a
 \right) ^{-1} \left( 4\,g-a \right) ^{-1} \\ &+2/3\, \left( 1+2\,{\frac {g
}{a}} \right) ^{2}z \left(  \left( 2\,{\frac {g}{4\,g-a}}+ \left( 2\,g
-a \right)  \left( u-{\frac {g+y}{z}} \right)  \left( 4\,g-a \right) ^
{-1} \right) ^{3}-{u}^{3} \right) \\ &+ \left( 1+2\,{\frac {g}{a}}
 \right) ^{2} \left( 2\,{\frac {g}{4\,g-a}}+ \left( 2\,g-a \right) 
 \left( u-{\frac {g+y}{z}} \right)  \left( 4\,g-a \right) ^{-1}
 \right) ^{2} \left( 2\,g-a \right)  \left( g+y \right)  \left( 4\,g-a
 \right) ^{-1} \\ &+ \left( 1+2\,{\frac {g}{a}} \right)  \left( g+y-z
 \left( u+2\,{\frac {g}{a}} \right)  \right)  \left(  \left( 2\,{
\frac {g}{4\,g-a}}+ \left( 2\,g-a \right)  \left( u-{\frac {g+y}{z}}
 \right)  \left( 4\,g-a \right) ^{-1} \right) ^{2}-{u}^{2} \right) \\ &+
 \left( 1+2\,{\frac {g}{a}} \right) z \left( -u-2\,{\frac {g}{a}}
 \right)  \left(  \left( 2\,{\frac {g}{4\,g-a}}+ \left( 2\,g-a
 \right)  \left( u-{\frac {g+y}{z}} \right)  \left( 4\,g-a \right) ^{-
1} \right) ^{2}-{u}^{2} \right) \\ & +2\, \left( 1+2\,{\frac {g}{a}}
 \right)  \left( g+y-z \left( u+2\,{\frac {g}{a}} \right)  \right) 
 \left( 2\,{\frac {g}{4\,g-a}}+ \left( 2\,g-a \right)  \left( u-{
\frac {g+y}{z}} \right)  \left( 4\,g-a \right) ^{-1} \right) \\
& \left( 2
\,g-a \right)  \left( g+y \right) {z}^{-1} \left( 4\,g-a \right) ^{-1}
\\ &+ \left( 2\,g-a \right)  \left( g+y \right)  \left( g+y-z \left( u+2\,
{\frac {g}{a}} \right)  \right) ^{2}{z}^{-2} \left( 4\,g-a \right) ^{-
1} \\ &+2\, \left( 2\,{\frac {g}{4\,g-a}}+ \left( 2\,g-a \right)  \left( u-
{\frac {g+y}{z}} \right)  \left( 4\,g-a \right) ^{-1}-u \right) 
 \left( g+y-z \left( u+2\,{\frac {g}{a}} \right)  \right)  \left( -u-2
\,{\frac {g}{a}} \right) 
\end{split}
\end{equation*}
\begin{verbatim}
 > latex(factor(%));
\end{verbatim}
\begin{equation*}
1/3\,{\frac { \left( 4\,uzg+ga+ay-2\,{g}^{2}-2\,gy-4\,gz \right) 
 \left( -2\,uzg+ga+ay-2\,{g}^{2}-2\,gy+2\,gz \right) ^{2}}{{z}^{2}
 \left( -4\,g+a \right) {a}^{2}}}
\end{equation*}
Maple is a trademark of Waterloo Maple Inc.
\end{appendices}
\end{document}